 \documentclass[12pt, draftclsnofoot, onecolumn]{IEEEtran} 
\usepackage{cite} 

\usepackage[T1]{fontenc}    
\usepackage[utf8]{inputenc} 
\usepackage{newtxtext,newtxmath} 
\usepackage[english]{babel} 

\usepackage{amsmath, amsfonts, mathtools} 

\usepackage{graphicx}
\usepackage[caption=false, font=footnotesize]{subfig} 
\usepackage{booktabs}  
\usepackage{array}     
\usepackage{booktabs}
\usepackage{caption}
\usepackage{float}
\usepackage{lscape}
\usepackage{multirow}
\usepackage{adjustbox}

\usepackage[ruled,vlined]{algorithm2e}  
\RestyleAlgo{ruled}  

\usepackage[hidelinks]{hyperref} 

\usepackage{textcomp}
\usepackage{verbatim}

\usepackage{stfloats} 

\usepackage{url} 

\addto\captionsenglish{}
\usepackage{balance}
\begin{document}

\title{Mitigating Phase Errors to Improve Signal Quality in RIS-Assisted Satellite Communications}

\author{M Khalil, \textit{Member, IEEE},  Ke Wang, \textit{ Member, IEEE}, Jiao Lin, \textit{Member, IEEE}, and Jinho Choi, \textit{Fellow, IEEE} 
  
}

\maketitle
\begin{abstract}
This research presents an advanced framework designed to enhance the received power in satellite-to-Earth communications by utilizing Reconfigurable Intelligent Surfaces (RIS) and focuses on mitigating phase errors arising from hardware imperfections associated with RIS systems. A comprehensive analysis of the phase errors arising from these imperfections is conducted, leading to the development of a robust analytical model that quantitatively incorporates these errors into the assessment of received power evaluations.  Subsequently, we propose a methodology to selectively exclude RIS elements that are prone to errors, thereby improving the phase alignment of the received signal and enhancing overall system efficiency. While this strategy leads to a marginal decrease in received power, Bayesian Optimization (BO) is employed to optimize the RIS configuration, maintaining the desired power levels and ensuring signal integrity. 
 The research also delves into the complexities introduced by shadowing effects combined with phase errors. To address these compounded challenges, a decision-making framework utilizing targeted BO is introduced to dynamically optimize RIS configurations, thereby enhancing system robustness and performance under adverse operational conditions.  Numerical simulations validate the framework's efficacy in adaptively managing RIS elements, ensuring robust signal integrity and improved reception despite variations stemming from environmental and hardware factors. By addressing critical challenges in RIS-augmented satellite communications, this work highlights the transformative potential of adaptive optimization strategies in advancing the reliability and efficiency of next-generation wireless networks. 
\end{abstract}
\section{Introduction }
Satellite communications are essential for modern wireless systems, offering expansive geographical coverage. With escalating demands for higher data rates and enhanced signal quality \cite{Zhao2024}, ongoing research has focused on advancing satellite communication links, particularly regarding signal integrity over long distances \cite{Du2023}, \cite{Ma2022}. For instance, \cite{Na2022} examines outage probability and symbol error rate in satellite systems with randomly distributed users, quantifying the impact of propagation loss and shadowed-Rician fading. Using stochastic geometry, the study derives asymptotic expressions to optimize high-power satellite operations. However, the vast distances between satellites and ground stations present significant challenges for system optimization \cite{Seyedi2013}. 

Reconfigurable Intelligent Surfaces (RIS) have emerged as a transformative solution by dynamically controlling the electromagnetic environment \cite{Ramezani2023} and \cite{Zhao2024a}. An RIS comprises cost-effective, reconfigurable elements that can be either passive or active \cite{Zhao2024b}. Passive RIS elements adjust the phase
of reflected signals without external power, thereby reducing energy consumption and hardware costs \cite{DiRenzo2020}, \cite{Basar2019}. In contrast, active RIS elements, while requiring minimal power, allow for real-time reconfiguration to enhance adaptability in dynamic environments \cite{Khalil2022}. 

The integration of RIS in satellite-to-Earth communications holds promise for addressing persistent challenges such as propagation loss, shadowed-Rician fading, and dynamic channel conditions that are inherent in long-distance transmissions \cite{Feldhake1997}. By intelligently reconfiguring signal reflection paths, RIS enhances received signal strength and compensates for atmospheric attenuation, ensuring robust connectivity in scenarios where traditional terrestrial infrastructure is impractical. This capability is particularly critical for satellite-dependent applications such as Vehicle-to-Everything (V2X) systems, which demand ultra-reliable and low-latency communication links to support mission-critical coordination between vehicles, infrastructure, and networks in environments with limited line-of-sight or ubiquitous coverage requirements.

Recent research has focused on optimizing the integration of RIS within satellite-terrestrial networks, addressing both performance benefits and operational challenges. For example, \cite{Niu2023} introduces an active RIS-assisted secure transmission scheme for cognitive satellite-terrestrial networks, achieving enhanced secrecy performance through the joint optimization of beamforming, artificial noise, and reflection coefficients, effectively mitigating dual fading. Similarly, \cite{Lv2024} investigates energy-efficient strategies in an RIS-assisted satellite-terrestrial network utilizing Non-Orthogonal Multiple Access (NOMA), employing advanced beamforming and interference management techniques. Moreover, \cite{Xu2021c} examines the role of RIS in Space-Air-Ground Integrated Networks (SAGIN) to improve connectivity and coverage, particularly in overcoming urban challenges posed by buildings and trees. This study demonstrates notable enhancements in coverage, interference suppression, and energy efficiency. Additionally, \cite{Ye2022} investigates RIS as a cost-effective solution in non-terrestrial communications to overcome blockages and dynamic environments, thereby extending coverage and improving signal quality across aerial and satellite systems. Furthermore, \cite{Khan2022} presents a sustainable optimization framework for RIS-assisted geostationary (GEO) satellite communications in 6G networks, utilizing a Mesh Adaptive Direct Search (MADS) algorithm to optimize satellite transmission power and RIS phase shifts, resulting in significant improvements in channel capacity compared to non-RIS baselines. 

Nonetheless, the real-world deployment of RIS faces significant challenges due to non-ideal environmental and hardware conditions. Beyond the shadowing effects caused by obstructions in satellite-to-ground paths \cite{Loo1985},\cite{Cavdar2003}, hardware-induced phase-shift errors critically degrade the performance of RIS-enabled systems. These errors arise from imperfections in RIS components, such as quantized phase resolution, manufacturing tolerances, and thermal noise in tunable elements like varactor diodes or PIN diodes. Such imperfections can cause deviations between intended and actual phase adjustments.  Even minor phase misalignments can accumulate across large-scale RIS arrays, reducing the beamforming gain and  the effective signal-to-noise ratio at the receiver \cite{Khalid2023}. In satellite communications, where precise phase coherence is critical to compensate for long-distance path loss and Doppler shifts, these errors exacerbate outage probabilities and compromise the reliability of critical applications such as real-time V2X coordination. To address these challenges and fully exploit RIS in satellite-terrestrial systems, this paper proposes a novel method designed to mitigate the adverse effects of RIS phase errors. The proposed method begins with an assessment of the impact of phase errors on the received signal and  the identification of problematic RIS elements, which are subsequently isolated to prevent cascading performance deterioration. While this isolation process inherently reduces the effective aperture (the effective area covered by the operational elements of the RIS)  of the RIS, an optimization procedure is introduced to enhance overall performance. This approach also accounts for shadowing effects in addition to RIS phase errors. Mathematical analysis and numerical simulations confirm that the proposed method improves received power by approximately 6.6\% to 9.2\% without shadowing and by about 9.3\% to 12.6\% when shadowing is considered. 

\subsection*{Related Works}
  Recent research has highlighted the transformative potential of RIS technology in enhancing coverage, capacity, energy efficiency, and interference suppression in wireless communication systems \cite{Wu2019a,Huang2019a,Zhou2020,Chen2021,Zhang2020a}. Most early studies concentrated on theoretical RIS phase shift optimizations under idealized RIS hardware conditions. Extending this theoretical foundation, \cite{Son2023} practical challenges have been addressed, focusing on phase shift designs aimed at mitigating inter-satellite interference in integrated networks. By employing passive beamforming, this study maximized the Signal-to-Interference-plus-Noise Ratio (SINR) for terrestrial users, thereby enhancing reliability and bridging the gap between theoretical models and real-world implementations.

At high frequencies, the performance of RIS is fundamentally constrained by hardware limitations (HL), which include phase inaccuracies, thermal distortions, and material nonlinearities. These imperfections can significantly degrade system efficiency and reliability, presenting a major barrier to the widespread deployment of RIS \cite{Yang2023}.

In addition,  the shorter wavelengths at high frequencies further amplify the effects of even minor hardware imperfections. For example, small thermal expansions in the RIS substrate materials (e.g., dielectric layers or metallic patches) can alter the effective electrical length of individual elements, leading to frequency-dependent phase offsets. These offsets reduce phase coherence, increase beamwidth, and diminish beamforming gain, thereby lowering the signal-to-noise ratio (SNR). Such degradation is particularly critical in narrow-beam satellite systems, where precise beam alignment is essential for maintaining robust link quality \cite{Trichopoulos2022}.

In addition to thermal distortions, prolonged exposure to high-frequency signals accelerates the aging of key RIS components, such as PIN diodes. Over time, this aging increases component resistance and introduces nonlinear phase deviations, which further degrade the performance of the RIS \cite{Wang2023}. 

 Frequency-dependent phase errors may also lead to beam squint shifts in the angle of reflected beams across frequency bands, causing inter-carrier interference in wideband orthogonal frequency-division multiplexing (OFDM) systems. This interference complicates signal detection and demodulation, impairing the overall communication link \cite{Wang2019a}. Moreover, dynamic misalignments driven by environmental factors or hardware instabilities introduce additional challenges, especially for applications that require real-time beam realignment, such as satellite-to-ground communications \cite{Hu2018,Alegria2019}.

To address these challenges, advanced modeling frameworks have been developed to quantify and mitigate HL in RIS-assisted systems. Gaussian process models, for instance, provide a statistical foundation to analyze how hardware imperfections degrade system capacity and increase interference in large intelligent surfaces, enabling precise characterization of performance bottlenecks \cite{Bjoernson2014,Azam2020}. Extensions of these methods evaluate transceiver HL in RIS-aided networks, revealing their adverse effects on secrecy performance (e.g., eavesdropping resilience) and spectral efficiency \cite{Boulogeorgos2020a,Chen2021c}. Such studies highlight the necessity of hardware-aware models that account for practical non-idealities, bridging the gap between theoretical performance bounds and real-world deployment constraints.

Reflecting this shift in research focus, recent work emphasizes its importance. For example, \cite{Chu2022} developed a robust framework for RIS-assisted integrated satellite-UAV-terrestrial networks, demonstrating that optimizing RIS phase shifts alongside UAV relay deployment mitigates outage probabilities caused by hardware impairments and interference. Similarly, \cite{Diao2024} designed a secure and energy-efficient UAV-based RIS system that is resilient to air turbulence and mechanical vibrations, leveraging UAV mobility to dynamically adjust signal paths when direct links are obstructed. These studies highlight how hardware-aware models enhance both reliability and adaptability in dynamic environments. Building on these insights, recent contributions  \cite{Nguyen2022}, \cite{Shang2022}, \cite{Liu2021}, \cite{Pan2021}, \cite{Hum2014} further emphasize the necessity of HL-aware design in RIS systems. These works integrate realistic imperfections, such as phase noise, quantization errors, and thermal drift, into optimization frameworks, enabling RIS to maintain coherent performance under practical operating conditions. Such advancements are critical for scalable, real-world deployments. However, while studies like \cite{Dai2022} and \cite{Khalid2022} model HL as random additive noise, their abstractions risk oversimplification. For instance, they often neglect to quantify how specific HL parameters (e.g., noise variance or phase-error thresholds) degrade signal quality or specify the tolerance thresholds required to guaranty target quality-of-service (QoS) metrics.

To bridge this gap, \cite{Pan2022a} systematically reviews robust signal processing techniques for RIS-aided systems, focusing on channel estimation and transmission designs that explicitly consider hardware nonidealities. Building on this, \cite{Badiu2020} proposes a statistical framework for modeling phase-shift errors using Von Mises distributions, revealing a critical trade-off: large RIS arrays utilize spatial averaging over Nakagami fading channels to mitigate phase errors, achieving near-ideal communication performance without the necessity for high-resolution phase control. In contrast, smaller arrays lack this spatial diversity and require precise phase alignment to prevent significant performance degradation. Despite these insights, existing works, such as \cite{Badiu2020}, \cite{Yang2023}, and \cite{Khalid2022}, remain largely theoretical. They primarily focus on the asymptotic behavior in large arrays, high-frequency channel vulnerabilities, or simplified impairment models. As a result, these studies provide limited practical strategies for addressing phase errors, which are crucial in high-precision and high-reliability applications such as satellite communications. This deficiency is particularly significant in light of the increasing interest in and utilization of high-frequency bands (e.g., X-band, S-band, and Ka-band \cite{Panagopoulos2004}) for deep-space missions, where hardware imperfections are magnified, and traditional terrestrial-focused mitigation approaches are insufficient.
\subsection*{Contributions}
Diverging from prior terrestrial studies, our work explicitly targets phase-error mitigation in ground-based RIS-assisted satellite links, addressing the unique challenges of high-frequency operations. We propose a dual-objective framework that strategically deactivates error-prone RIS elements while actively optimizing the radiation pattern of the remaining units through Bayesian optimization (BO). Although deactivation reduces the RIS aperture, theoretically decreasing the received power, BO dynamically reconfigures the operational elements to compensate for these aperture losses. This strategy ensures robust performance, even in the presence of the heavy-tailed phase-error distributions identified in \cite{Dai2022} and \cite{Khalid2022}. This adaptive approach effectively mitigates errors while preserving signal integrity, representing a significant advancement for satellite systems where static RIS configurations are insufficient.

To the best of our knowledge, this study presents the first systematic demonstration of performance-aware element exclusion for phase error mitigation in satellite links. Simulations conducted under a practical additive error model confirm that our approach enhances received power by approximately 6.6\% to 9.2\%. Moreover, when considering shadowing effects, the improvement in received power increases to a range of 9.3\% to 12.6\%. These results highlight the effectiveness of selective deactivation as a cost-effective strategy to balance hardware complexity, aperture size, and throughput in high-frequency regimes. By aligning hardware-aware optimization with the specific constraints of satellite-terrestrial channels, such as Doppler shifts, atmospheric attenuation, and dynamic misalignment, our work advances the feasibility of RIS in next-generation space-ground networks.

This study introduces a rigorously designed framework aimed at addressing critical challenges in RIS-assisted satellite-to-Earth communications. Specifically, the framework targets performance degradation caused by hardware imperfections in RIS elements and environmental shadowing. These imperfections include manufacturing flaws, environmental factors that lead to material degradation, and operational wear and tear, which can significantly distort the intended phase of reflected signals \cite{Yang2023}. Our approach is distinctive in emphasizing these practical hardware challenges, particularly those significant at the high frequencies utilized in satellite communications. At these frequencies, even minor imperfections can have substantial impacts. The framework addresses these issues through the following key contributions:
\begin{itemize}
\item We systematically characterize the origins of phase errors in RIS hardware, specifically focusing on high-frequency satellite applications where the inherent sensitivity to imperfections is exacerbated by shorter wavelengths. Building on this analysis, we develop a comprehensive received power model that explicitly incorporates hardware-induced phase errors. The model establishes a closed-form relationship between phase inaccuracies and received power degradation, enabling direct evaluation of performance losses caused by imperfect RIS elements. By quantifying these practical challenges, the framework also enables further optimization of RIS configurations and conducting sensitivity analyzes under real-world hardware constraints.
\item To address phase errors in impaired RIS elements, we propose a threshold-based deactivation method. RIS elements exhibiting phase errors that exceed a predefined threshold are selectively disabled,
thereby preventing excessive destructive interference, which diverges from conventional methods that assume uniform element performance. To mitigate reductions in received power resulting from the exclusion of certain elements, we employ a BO algorithm. This algorithm simultaneously optimizes the phase shifts of the remaining (non-excluded) RIS elements and dynamically adjusts the error threshold. Our simulation results demonstrate that the BO algorithm significantly enhances received power and SINR under both additive and multiplicative error models. 
\item The framework is further extended to address realistic urban environments where satellite signals are attenuated by shadowing from buildings, foliage, and terrain. These shadowing effects, statistically modeled using log-normal distributions, exacerbate the performance degradation caused by hardware-induced phase errors. To mitigate these dual impairments, we develop a Monte Carlo--based procedure that accounts for stochastic shadowing and phase misalignments. This method adaptively identifies and mitigates problematic RIS elements under practical conditions. Comprehensive simulation-based evaluations confirm that our combined threshold-based exclusion and Bayesian optimization approach consistently maintains signal integrity and enhances received power, effectively countering the challenges posed by both shadowing and hardware imperfections. The system\textquoteright s robustness is numerically validated under dynamically varying environmental conditions, demonstrating the viability of our adaptive control strategy for RIS elements in realistic scenarios.
\end{itemize}
 To ensure clarity and consistency in the analysis of our system model, we summarize the key notations and definitions
in Table~\ref{tab:notation}. 

\begin{table}[ht]
\centering
\caption{ Notation and Definitions}
\label{tab:notation}
\renewcommand{\arraystretch}{1.1} 
\setlength{\tabcolsep}{5pt}       
\resizebox{\columnwidth}{!}{%
\begin{tabular}{|c|l|}
\hline
\textbf{Symbol} & \textbf{Definition} \\ \hline
$U$ & User terminal \\ \hline
$d_{\mathrm{SR}}$ & Distance from source (S) to RIS \\ \hline
$d_{\mathrm{SU}}$ & Distance from source (S) to user (U) \\ \hline
$d_{\mathrm{RU}}$ & Distance from RIS to user (U) \\ \hline
$L_{\mathrm{SU}}$ & Path loss between S and U \\ \hline
$L_{\mathrm{SR}}$ & Total path loss between S and RIS \\ \hline
$L_{\mathrm{RU}}$ & RIS-to-user path loss \\ \hline
$N$ & Total number of RIS elements \\ \hline
$\Gamma_{n}$ & Reflection coefficient of RIS element $n$ \\ \hline
$\left|\Gamma_{R}\right|$ & Effective amplitude reflection coefficient \\ \hline
$\Theta$ & Total phase difference \\ \hline
$\phi_{n}$ & RIS phase shift \\ \hline
$\phi_{R}$ & Collective phase shift \\ \hline
$\delta$ & Error factor \\ \hline
$\sigma_{th}$ & Phase error threshold \\ \hline
$\overset{\hookrightarrow}{\theta}$, $\theta$ & Aggregated incident and reflection angles \\ \hline

$x_{\mathrm{RU}}(\theta_{u})$ & Terrestrial shadowing \\ \hline
$\eta_{\mathrm{SR}}(\overset{\hookrightarrow}{\theta_{n}})$ & Additional path loss between S and RIS \\ \hline
$\eta_{\mathrm{SU}}(\theta_{u})$ & Elevation-angle-dependent clutter/shadowing loss \\ \hline
$\xi$ & Exploration-exploitation trade-off parameter \\ \hline
$P_{R}$ & Received power \\ \hline
UCB, EI & Upper confidence bound and expected improvement criteria \\ \hline
\end{tabular}
}
\end{table}
The remainder of this paper is structured as follows: Section \ref{sec:Syst_Mod} outlines the system model, detailing the architecture and operational principles of the RIS-assisted satellite-to-Earth communication framework. Section \ref{P-non-ideal} rigorously analyzes the degradation of received power caused by hardware-induced phase shift errors, quantifying their impact on system performance. Section \ref{sec:Opt-non_idel} presents a robust optimization framework designed to mitigate phase uncertainties, thereby enhancing the quality of the received signal through adaptive RIS reconfiguration. Section \ref{sec:non_idel} extends the analysis to incorporate shadowing effects and develops a joint optimization and compensation strategy for practical urban environments. Finally, Section \ref{Conclusion} summarizes the key findings, discusses practical implications, and outlines future research directions.
\section{System Model\label{sec:Syst_Mod} }
\begin{figure}
\begin{centering}
\includegraphics[width=3.5in,viewport=2bp 0bp 550bp 350bp]{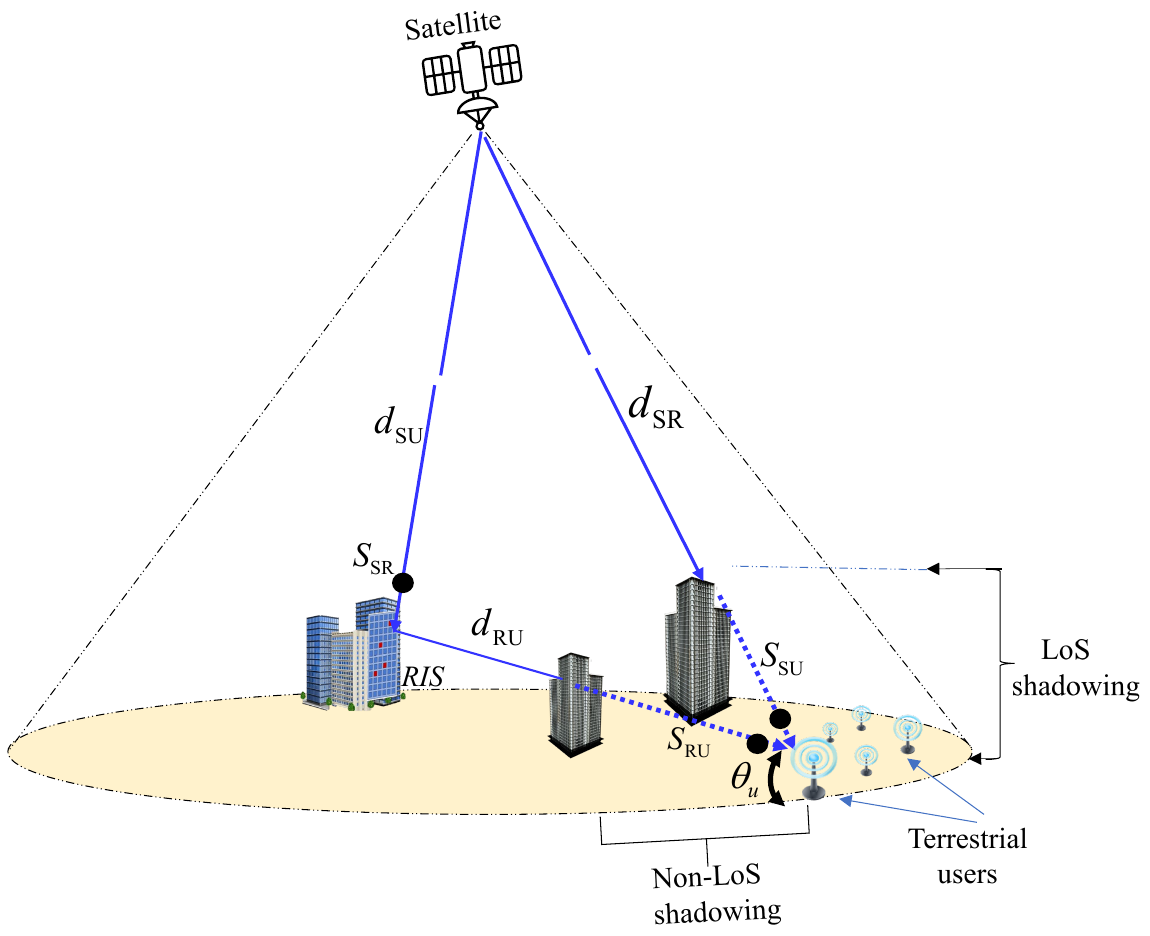} 
\par\end{centering}
\caption{Overview of RIS-aided satellite communication system\label{fig:0}}
\end{figure}

This section outlines the architecture of the examined satellite-to-ground communication system. The proposed framework incorporates a terrestrial RIS to relay signals between the satellite (S) and the user terminal (U), establishing dual transmission pathways: an indirect RIS-assisted link and a direct line-of-sight (LOS) channel. Fig. \ref{fig:0}, illustrates the geometric arrangement and spatial configuration of S, RIS, and U. To specifically examine the impact of hardware-induced phase errors on RIS performance, we analyze a simplified single-user scenario. This setup minimizes other influences, such as multi-user interference and environmental variations, enabling a focused evaluation of phase error effects. This approach also provides a solid foundation for future expansion to multi-user systems. Furthermore, using a single scalar parameter simplifies our secondary goal of optimizing the RIS phase shifts, streamlining the design process and increasing adaptability for a wider range of applications.

The distances involved in signal propagation are defined as $d_{\mathrm{SR}}$ (from S to RIS), $d_{\mathrm{SU}}$ (from S to U directly), and $d_{\mathrm{RU}}$ (from RIS to U). Given the global scale of satellite communications and the spherical geometry of the Earth, spherical trigonometry is employed to calculate these distances effectively. The elevation angles from the satellite to the RIS and the user terminal are denoted by $\theta_{sr}$ and $\theta_{su}$(in radians), respectively. These angles are determined based on the known geographic coordinates, specifically the latitude and longitude, of the satellite, RIS, and user terminal \cite{Vallado2001}. 

Doppler shifts resulting from the satellite's high velocity relative to the ground receiver are compensated for by advanced signal processing algorithms implemented in the user terminal. Furthermore, atmospheric effects, including ionospheric delay and tropospheric attenuation, are effectively mitigated through the use of phased-array beam steering and adaptive modulation techniques. These assumptions are consistent with the foundational principles of contemporary satellite communication systems.

Assuming the RIS is located at coordinates $\bigl(\text{lat}_{\mathit{RIS}},\text{lon}_{\mathit{RIS}}\bigr)$ and the user terminal U is located at $\bigl(\text{lat}_{u},\text{lon}_{u}\bigr)$, the geodesic distance between the RIS and U, denoted as $d_{\mathrm{RU}}$, is computed using the Vincenty formula. This formula calculates the shortest path (i.e., the geodesic) between two points on an ellipsoidal model of the Earth, offering higher accuracy than spherical approximations, especially over long distances. For further mathematical details, please refer to Vincenty's original work \cite{Mahmoud2016}. The geodesic distance is calculated as follows:
\begin{equation}
d_{Rk}=\mathrm{\textrm{VincentyDistance}}(\mathrm{\text{lat}_{\mathit{RIS}},\text{lon}_{\mathit{RIS}},\text{lat}_{u},\text{lon}_{u}})
\end{equation}
Here, $\mathit{\mathrm{lat}}$ and $\mathit{\mathrm{lon}}$ denote the latitude and longitude of the respective points provided by GPS.

To determine the total received power at user U from both the LOS  path and the reflected path via the  RIS, we begin by calculating the power of the direct LOS path from the satellite to U. The complex signal arriving at user U, denoted $S_{\mathrm{SU}}$, is derived using the Friis transmission equation, as follows:
\begin{equation}
S_{\mathrm{SU}}=\sqrt{\frac{P_{\mathrm{t}}G_{\mathrm{t}}G_{\mathrm{U}}}{L_{SU}}}e^{-j\phi_{\mathrm{SU}}},\label{eq: Ssu}
\end{equation}
where $P_{\mathrm{t}}$ is the transmitted power, $G_{\mathrm{t}}$
and $G_{\mathrm{U}}$ are the transmitter and receiver antenna gains, respectively, $\phi_{\mathrm{SU}}$ is the phase of the directly received
signal, which equals $\frac{2\pi}{\lambda}\mathit{d_{\mathrm{SU}}}$
with $\lambda$ representing the signal wavelength, and $L_{\mathrm{SU}}$
is the path loss between S and U. The path loss is primarily attributed
to free-space propagation and is expressed as:
\begin{equation}
L_{\mathrm{SU}}\left[\mathrm{dB}\right]=L_{FS}(d_{\mathrm{SU}},f)+L_{\tau}(\theta_{u},\mathit{f_{c}})+\eta_{\mathrm{SU}}(\theta_{u}),\label{eq: LSU}
\end{equation}
where $L_{FS}$ is the free-space path loss calculated as $20\log_{10}(f_{\mathrm{c}})+20\log_{10}(d)-147.55,$$L_{\tau}(\theta_{u},\mathit{f_{c}})$
represents atmospheric absorption loss, and $\eta_{\mathrm{SU}}(\theta_{u})$
is an elevation-angle-dependent clutter/shadowing loss defined by
the 3GPP model \cite{3GPP2019}, encompassing building obstructions
and log-normal shadow fading. It is important to note that the elevation
angle $\theta_{u}$, which defines the signal angle between the RIS
and the user, is assumed to be constant in this model.

The path of a signal reaching the RIS can be expressed as 
\begin{equation}
\overset{\hookrightarrow}{S_{SR}}=\sqrt{\frac{P_{t}G_{\mathrm{t}}\mathrm{\mathrm{G(\overset{\hookrightarrow}{\theta_{n}})}}}{L_{\mathrm{SR}}}}e^{-j\phi_{\mathrm{SR}}},
\end{equation}
where $G(\overset{\hookrightarrow}{\theta_{n}})$ is the antenna gain
at the $\mathit{n}$$^{th}$ RIS element for the incidence angle $\overset{\hookrightarrow}{\theta_{n}}$,
and $L_{\mathrm{SR}}$ is the total path loss between S and RIS, expressed
in dB as: 
\begin{equation}
L_{\mathrm{SR}}\left[\mathrm{dB}\right]=L_{FS}(d_{\mathrm{SR}},f_{\mathrm{c}})+L_{\tau}(\overset{\hookrightarrow}{\theta_{n}},f_{\mathrm{c}})+\eta_{\mathrm{SR}}(\overset{\hookrightarrow}{\theta_{n}})
\end{equation}
where $\eta_{\mathrm{SR}}(\overset{\hookrightarrow}{\theta_{n}})$ is an additional path loss defined between S and R with respect to the incident angle. 

After interaction with the RIS, the reflected signal is:
\begin{equation}
S_{\mathrm{SR}}=\sqrt{\frac{P_{t}G_{\mathrm{t}}}{L_{\mathrm{SR}}}}e^{-j\phi_{\mathrm{SR}}}\sum_{n=1}^{N}\sqrt{G(\overset{\hookrightarrow}{\theta_{n}})G(\theta_{n})|\Gamma_{n}|^{2}}e^{-j\phi_{n}}.
\end{equation}
Here, $N$ represents the total number of RIS elements, $\Gamma_{n}$
represents the reflection coefficient of the $\mathit{n}$ RIS element, and $\phi_{n}$ denotes the phase shift introduced by the
$\mathit{n}^{th}$ RIS element.

The reflection coefficient $\Gamma_{n}$ represents the fraction of incident signal power reflected by the $\mathit{n}$$^{th}$ element of the RIS. For a passive RIS, $\Gamma_{n}$ is a real value ranging from 0 to 1, where 0 corresponds to no reflection and 1 to total reflection. Each RIS element independently adjusts the phase of the incident signal, enabling precise control over signal propagation. Assuming the RIS is typically placed to minimize unintended multipath signals, we focus exclusively on the RIS-routed signal while neglecting environmental multipath interference. Unlike random multipath effects, the RIS-assisted path complements the direct LOS path by carefully adjusting the signal's phase and direction, thereby improving overall system performance. Consequently, the signal from the RIS to the receiver, U, can be expressed as:
\begin{equation}
S_{\mathrm{RU}}=\sum_{n=1}^{N}\sqrt{\frac{P_{t}G(\overrightarrow{\theta_{n}})G(\theta_{n})G_{\mathrm{t}}}{L_{\mathrm{SR}}}}\left|\Gamma_{n}\right|\sqrt{\frac{G_{\mathrm{u}}}{L_{\mathrm{RU}}}}e^{-j(\phi_{\mathrm{RU}}+\phi_{\mathrm{SR}}+\phi_{n})},\label{eq:a}
\end{equation}
where $L_{\mathrm{RU}}$ is the RIS-to-U path loss in dB:
\begin{equation}
L_{\mathrm{RU}}\left[\mathrm{dB}\right]=L_{\mathit{FS}}(d_{\mathrm{RU}},f_{\mathrm{c}})+x_{\mathrm{RU}}(\theta_{u}).\label{eq:LRU}
\end{equation}
Here, $x_{\mathrm{RU}}(\theta_{u})$ represents the terrestrial shadowing component caused by clutter-induced random fluctuations, modeled as
a log-normal distribution $x_{\mathrm{RU}}\sim\mathrm{\log}(\mu_{\mathrm{ru}},\sigma_{\mathrm{ru}})$, with $\mu_{\mathrm{ru}}$ and $\sigma_{\mathrm{ru}}$ as the mean and variance in dB. The assumption of negligible multipath is justified by the RIS's strategic placement on a high building, which ensures a clear LOS to the receiver. Furthermore, higher operating frequencies and the RIS's design for directed signal transmission further reduce multipath effects and enhance efficiency.

The vector summation $\sum_{n=1}^{N}\Gamma_{n}e^{-j\phi_{n}}$ aggregates the amplitude and phase contributions of all RIS elements. For simplicity, this is reduced to two components: $\left|\Gamma_{R}\right|$ (effective amplitude reflection coefficient) and e$^{-j\phi_{R}}$ (collective phase shift). These terms capture the RIS's net impact on the signal's magnitude and phase. The revised expression for $S_{\mathrm{RU}}$ becomes:
\begin{equation}
S_{\mathrm{RU}}=\,\sqrt{\frac{P_{t}G_{\mathrm{t}}G_{\mathrm{U}}G(\overset{\hookrightarrow}{\theta})G(\theta)}{L_{\mathrm{SR}}L_{\mathrm{RU}}}}\left|\Gamma_{\mathrm{R}}\right|e^{-j\left(\phi_{\mathrm{RU}}+\phi_{\mathrm{SR}}+\phi_{\mathrm{\mathit{R}}}\right)},\label{eq:b}
\end{equation}
where $\overset{\hookrightarrow}{\theta}$ and $\theta$ represent the aggregated incident and reflection angles, respectively, simplifying
the net effect of the RIS's.

The total received power is the sum of contributions from both the direct and reflected paths. To quantify these contributions, the complex amplitudes for both paths are defined as follows: the direct path amplitude is given by
\begin{equation}
A_{\mathrm{SU}}=P_{t}\sqrt{\frac{G_{\mathrm{u}}G_{\mathrm{t}}}{L_{\mathrm{SU}}}},
\end{equation}  
and for the reflected path it is  
\begin{equation}
A_{\mathrm{RU}}=P_{t}\sqrt{\frac{G_{\mathrm{t}}}{L_{\mathrm{SR}}}}\sqrt{G(\overset{\hookrightarrow}{\theta})G(\theta)}\left|\Gamma_{\mathrm{R}}\right|\sqrt{\frac{G_{\mathrm{u}}}{L_{\mathrm{RU}}}}.
\end{equation}
The total received signal can then be expressed as follows: 
\begin{gather}
P_{\mathrm{R}}=\,\left|A_{\mathrm{SU}}e^{\phi_{1}}+A_{\mathrm{RU}}e^{\phi_{2}}\right|^{2}\nonumber \\
=P_{t}\left(A_{\mathrm{SU}}^{2}+A_{\mathrm{RU}}^{2}+2A_{\mathrm{SU}}A_{\mathrm{RU}}\cos(\phi_{1}-\phi_{2})\right),\label{eq: PR(cos)}
\end{gather}
where $\phi_{1}=-\phi_{\mathrm{SU}}$ and $\phi_{2}=-\left(\phi_{\mathrm{RU}}+\phi_{\mathrm{SR}}+\phi_{\mathrm{R}}\right).$
Rearranging $\phi_{1}$ and $\phi_{2}$ , we obtain:
\begin{equation}
P_{R}(\phi_{\mathrm{R}})=P_{t}\left[A_{\mathrm{SU}}^{2}+A_{\mathrm{RU}}^{2}+2A_{\mathrm{SU}}A_{\mathrm{RU}}\cos(\phi_{\mathrm{\mathit{R}}}-\Theta)\right]
\end{equation}  
where $\Theta=\phi_{\mathrm{SU}}-\phi_{\mathrm{SR}}-\phi_{\mathrm{RU}},$ represents the total phase difference resulting from the propagation paths in the system. It serves as a reference for adjusting the RIS phase to enhance signal alignment and improve system performance.

\section{Received Signal Power Analysis with phase shift error \label{P-non-ideal}}
Existing research on RIS optimization, such as studies by \cite{Wu2020} and \cite{Lin2022}, commonly assumes ideal RIS hardware conditions,
focusing solely on phase shift design without accounting for practical imperfections. In real-world deployments, RIS hardware experiences multiple non-idealities, including quantization limitations, hardware imperfections, calibration inaccuracies, and dynamic channel variations \cite{Chu2022}. As emphasized by \cite{Yang2023}, these practical issues introduce
phase errors, defined as unintended deviations in the RIS phase response due to factors such as temperature fluctuations, component aging, and manufacturing inconsistencies. These errors critically degrade performance, particularly in high-frequency satellite communication systems (e.g., Ka-band), where precise phase alignment is vital for maintaining signal coherence and link reliability. At such frequencies, shorter wavelengths
and increased power densities significantly amplify the system's sensitivity to minor physical deviations, resulting in substantial phase distortions caused even by minor hardware inaccuracies, such as material deformation or dielectric variation.

Additionally, high power densities cause increased thermal and mechanical stresses on RIS elements, leading to non-uniform temperature gradients and differential thermal expansion. These thermal and mechanical impacts collectively degrade the capability of the RIS to maintain its designated phase profile, thereby directly diminishing coherent signal gain. Although other sources of error, including quantization limits and channel estimation inaccuracies, remain relevant, the unique susceptibility of RIS-aided satellite systems to hardware-induced phase errors necessitates dedicated attention. Thus, this research specifically addresses practical methodologies for RIS design, calibration, and real-time operation under realistic hardware constraints, filling a significant gap in the existing literature, which often presumes idealized
conditions.
\subsection{Approaches for Modeling Phase Errors}
To accurately model RIS phase errors in satellite communication systems, we analyze their impact on the phase term $\bigl(\phi_{\mathrm{R}}-\Theta\bigr)$ appearing in \eqref{eq: PR(cos)} by introducing a random error factor. Two principal modeling approaches can be employed to describe the effects of this phase error:
\textit{Additive error model:} In this approach, the phase error $\delta,$ is directly added to the ideal phase term, yielding $(\phi_{\mathrm{R}}-\Theta)+\delta.$. This approach provides conceptual clarity and effectively represents various practical scenarios \cite{LorcaHernando2023}. The error can be characterized using multiple probability distributions based on specific hardware imperfections. A uniform distribution is suitable when the phase error is equally probable within a defined range, while a Gaussian distribution applies to scenarios where errors result from numerous small, independent disturbances (e.g., thermal noise). Due to its simplicity and analytical tractability, the additive error model facilitates straightforward error characterization and effective initial mitigation strategies.

\textit{Multiplicative error:} This model treats the error as a proportional scaling factor, described as $(\phi_{\mathrm{R}}-\Theta)$ $(1+\delta)$. Such a model accurately captures conditions where the error magnitude scales proportionally with the original phase value, exemplified by temperature-induced material deformations or power supply fluctuations \cite{Harger1967}. The multiplicative model effectively represents scenarios where external factors (e.g., temperature or voltage variations) exponentially influence phase shifts. Here, the error, $\delta,$ may follow a uniform distribution within bounded limits or a log-normal distribution if it arises from multiple multiplicative factors, which is commonly observed in complex multipath scenarios. Despite providing detailed insights into error propagation under dynamic conditions, the complexity of the multiplicative model can limit practical implementation, particularly when computational simplicity and real-time adaptability are essential.

Given the balance between analytical simplicity and practical fidelity to realistic hardware constraints, the additive error model is predominantly adopted in wireless systems research \cite{Tang2021a}. Therefore, our analysis employs an additive error framework, where the random error, $\delta,$ is uniformly distributed within the interval $[-\sigma,\sigma]$, expressed as $\delta\sim\mathcal{U}(-\sigma,\sigma)$. Here, $\sigma$ defines the maximum absolute phase error allowed by the system's operational requirements.

\subsection{Uniform Distribution of Phase Errors}
The adoption of a uniform distribution to model phase errors in RIS hardware is justified by its physical and operational constraints. Specifically, quantization limitations arising from the finite resolution of phase shifts in RIS elements restrict deviations to a well-defined interval, making all values within equally probable \cite{Chapala2023}. This bounded range is particularly suitable for modeling realistic RIS imperfections, which arise from subtle factors such as calibration tolerances, material inhomogeneities, and manufacturing inconsistencies \cite{Yang2023}.

Employing a uniform distribution provides several advantages. It inherently
captures the bounded nature of hardware-induced deviations, thus avoiding
unrealistic extreme errors associated with unbounded distributions
such as Gaussian distributions. Additionally, the uniform distribution's
constant probability density function (PDF) enhances analytical tractability,
facilitating its incorporation into stochastic signal models and allowing
for the derivation of closed-form performance metrics \cite{Bian2023}.
Finally, it naturally aligns with practical RIS operational limits,
where phase shifts must remain within defined boundaries, thereby
supporting realistic and robust performance evaluations.

\begin{figure}
\centering
\includegraphics[width=3.5in]{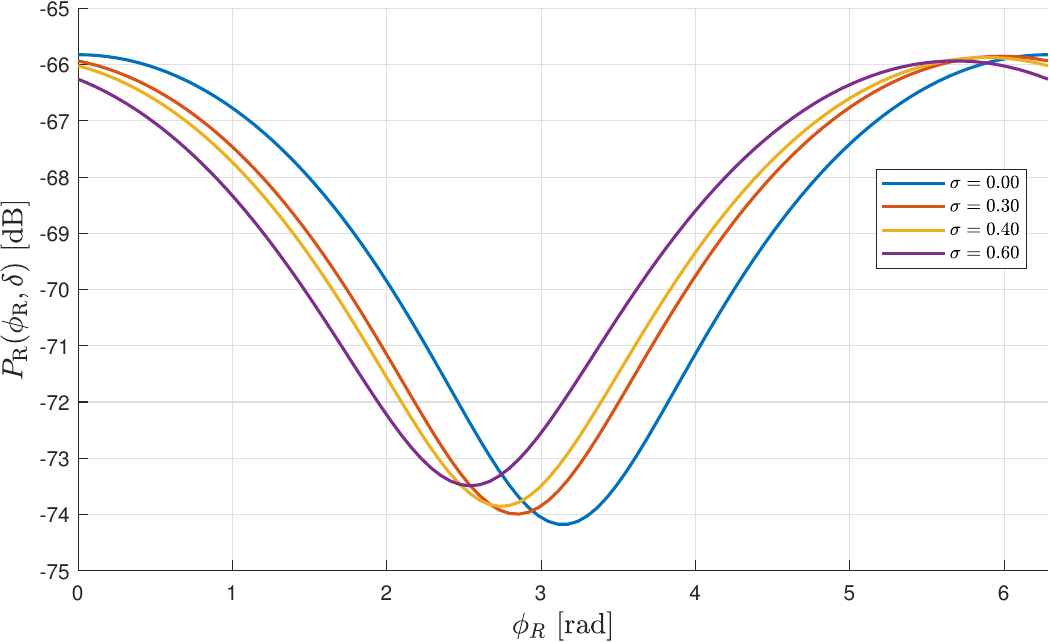}
\caption{Effect of phase-error magnitude on the expected received power in terms of $\phi_R$ and $\sigma$.}
\label{fig:1}
\end{figure}

Mathematically, the PDF of the uniformly distributed phase error is given by:
\begin{equation}
f(\delta)=\begin{cases}
\frac{1}{2\sigma}, & \text{if }-\sigma\leq\delta\leq\sigma\\
0, & \text{otherwise}
\end{cases}
\end{equation}
To explicitly analyze the instantaneous impact of random phase errors
on received power, the received power at the user terminal is expressed as a random function of the RIS phase shift and the instantaneous phase error:
\begin{equation}
P_{\mathrm{R}}\left(\phi_{\mathrm{R}},\delta\right)=P_{\mathrm{t}}\left(A_{\mathrm{SU}}^{2}+A_{\mathrm{RU}}^{2}+2A_{\mathrm{SU}}A_{\mathrm{RU}}\:\left[\cos\left(\phi_{\mathrm{R}}-\Theta+\delta\right)\right]\right).\label{eq:PR-B}
\end{equation}
$\;$ Fig. \ref{fig:1}, generated using the parameters outlined in Table \ref{tab:Table-1}, illustrates representative realizations of this instantaneous received power $P_{\mathrm{R}}\left(\phi_{\mathrm{R}},\delta\right)$  plotted against the RIS phase shift for several fixed values of the phase-error bound $\sigma$ . Each curve in Fig. \ref{fig:1} represents a single realization, demonstrating how instantaneous random errors affect the behavior of received power as $\phi_{\mathrm{R}}$ varies from $\phi_{\mathrm{R}}$$\approx0$ to $2\pi.$ In the ideal scenario($\sigma=0$), RIS elements are perfectly phase-aligned, resulting in a prominent constructive peak occurring near $\phi_{\mathrm{R}}$$\approx0$ or $2\pi,$ and a pronounced deep null near $\phi_{\mathrm{R}}\approx\pi$, reflecting maximum destructive interference.

 As the phase error bound increases, random instantaneous phase deviations shift and distort the received power pattern. This randomness leads to fluctuations and shifts in the locations of constructive peaks and destructive nulls for each individual realization. Consequently, peak power levels generally decrease due to imperfect coherent summation, while previously pronounced destructive nulls become less consistent and less deep, reflecting disrupted destructive interference. Notably, in the instantaneous realizations depicted, increasing visibility manifests as both a reduction in peak power (the upper regions shift downward) and elevated power levels in regions previously exhibiting deep destructive interference (the lower regions shift upward). Furthermore, due to randomness, there are noticeable lateral shifts (phase shifts) in the locations of maxima and minima across different realizations, clearly illustrating how instantaneous random errors influence real-time RIS-assisted signal coherence.
\begin{table} 
\centering
\renewcommand{\arraystretch}{1.3} 
\setlength{\tabcolsep}{10pt} 
\caption{Parameters Utilized for Various Example Illustrations}
\label{tab:Table-1}
\begin{tabular}{!{\vrule width 1.2pt} c !{\vrule width 1.2pt} c !{\vrule width 1.2pt} c !{\vrule width 1.2pt}}
\specialrule{1.2pt}{0pt}{0pt}  
\textbf{Parameter} & \textbf{Value} & \textbf{Range} \\
\specialrule{1.2pt}{0pt}{0pt}  
Transmit Power ($P_t$) & $10$ W & -- \\
Carrier Frequency ($f_c$) & $3$ GHz & -- \\
RIS Elements ($N$) & -- & $100 \leq N \leq 500$ \\
\specialrule{1.2pt}{0pt}{0pt}  
Distances ($d_{\mathrm{SU}}, d_{\mathrm{SR}}$) & 400 km & $0 \leq d_{\mathrm{SR}} \leq 1000$ m \\
Shadowing ($\sigma_{\mathrm{sh}}$) & -- & $0 \leq \sigma_{\mathrm{sh}} \leq 12$ dB \\
Phase Error ($\sigma$) & -- & $0 \leq \sigma \leq 0.5$ radians \\
\specialrule{1.2pt}{0pt}{0pt}  
Samples ($K$) & $10^5$ & -- \\
\specialrule{1.2pt}{0pt}{0pt}  
\end{tabular}
\end{table}
\section{RIS Optimization with Phase Error \label{sec:Opt-non_idel}}
Building on the analysis of non-ideal RIS-assisted satellite systems in Section \ref{P-non-ideal}, which identified significant degradation in received power due to instantaneous random phase errors, this section develops a systematic framework for evaluating and mitigating these errors to ensure reliable signal integrity. To isolate the direct impact of phase misalignment, we initially exclude other environmental factors, such as shadowing, thus enabling a focused investigation into effective phase-error correction strategies. This provides a foundational methodology for subsequent extensions that incorporate broader channel impairments.
\subsection{RIS Element Reliability Modeling under Phase Error}
The operational reliability of each RIS element is assessed using a defined phase-error threshold $\sigma_{th}$ that specifies the maximum allowable deviation from the intended phase configuration. RIS elements exceeding this threshold are classified as error-prone $\bigl(\mathcal{S_{\mathit{i}}}=1\bigr)$ and are dynamically excluded from signal reflection. Conversely, elements within the threshold retain full operational status $\bigl(\mathcal{S_{\mathit{i}}}=0\bigr)$.
The reliability status of the $i^{th}$ RIS element is mathematically defined as:
\begin{equation}
\textrm{\ensuremath{\mathcal{S_{\mathit{i}}}}}\left(\delta_{i},\sigma_{th}\right)=\left\{ \begin{array}{cc}
0 & \quad\text{if }\left|\delta_{i}\right|\leq\sigma_{th}\quad\left(\textrm{ Error-free}\right)\\
1 & \quad\textrm{otherwise\text{ \quad\ensuremath{\left(\textrm{ \textrm{Error-prone}}\right)}.}}
\end{array}\right.
\end{equation}
where $\delta_{i}$ is the instantaneous phase error of the $i^{th}$ element, uniformly distributed as $\delta\sim\mathcal{U}(-\sigma,\sigma)$, with the error's half-width $\sigma,$ representing the bound on phase errors due to realistic RIS imperfections. This distribution captures bounded hardware imperfections inherent in practical RIS deployments, such as quantization limits or calibration inaccuracies.
 The threshold $\sigma_{th},$ is chosen within $[0,\pi]$ radians, covering scenarios from perfect alignment (no tolerance for error) to the maximum practically tolerable deviation.

The instantaneous power at the user terminal, incorporating selective phase-error exclusion and accounting for both direct and RIS-reflected paths, is expressed as:
\begin{gather}
P_{\mathrm{R}}\left(\phi_{\mathrm{R}},\sigma_{th}\right)=P_{\mathrm{t}}\left(A_{\mathrm{SU}}^{2}+\sum_{i=1}^{N}\textrm{\ensuremath{(1-\mathcal{S_{\mathit{i}}}}})\:A_{\mathrm{RU_{\mathit{i}}}}^{2}\right.\nonumber \\
\left.+2A_{\mathrm{SU}}\sum_{i=1}^{N}\textrm{\ensuremath{(1-\mathcal{S_{\mathit{i}}}}})A_{\mathrm{RU_{\mathit{i}}}}^{2}\:\cos\left(\phi_{\mathrm{R}}-\Theta+\delta_{i}\right)\right),\label{eq: Test}
\end{gather}
In Eq. \eqref{eq: Test}, the term \textbf{$\bigl(1-\mathcal{S_{\mathit{i}}}\bigr)$} ensures that only RIS elements whose instantaneous phase errors satisfy $\left|\delta_{i}\right|\leq\sigma_{th}$ contribute to the coherent superposition. The cosine term explicitly accounts for residual phase errors from these elements. Selectively excluding elements with significant phase misalignment prevents destructive interference and enhances the overall robustness of the RIS-assisted communication link.

\begin{figure}
\centering{}
\includegraphics[width=3.5in]{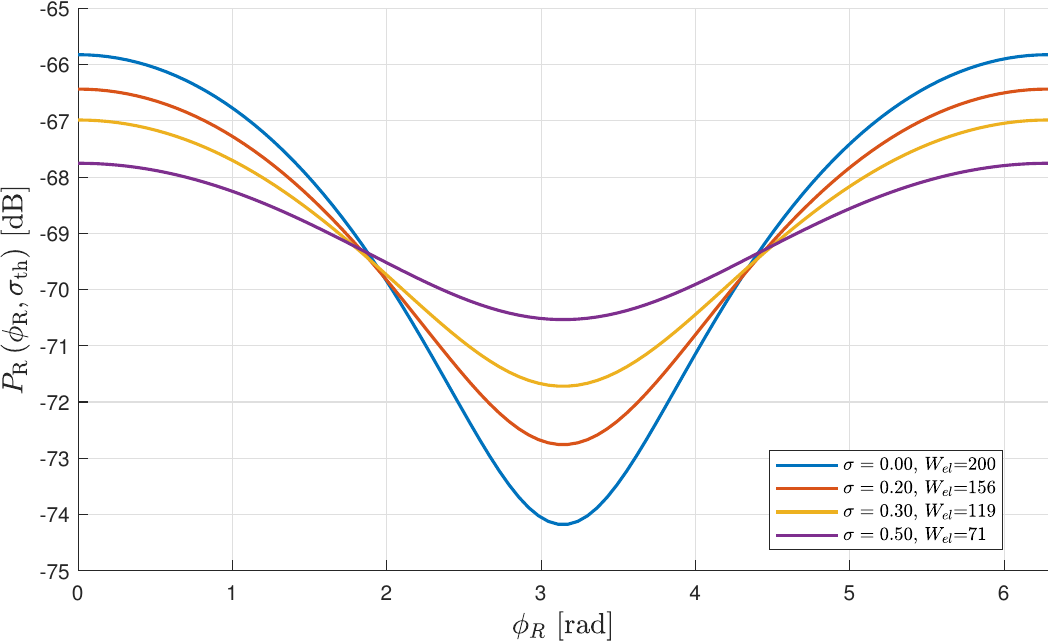}
\caption{Optimizing received power by selecting working RIS elements ($W_{el}$)based on phase adjustment\label{fig 2} }
\end{figure}

Fig. \ref{fig 2} illustrates the selective-exclusion approach defined by Eq. \eqref{eq: Test}. Under this approach, RIS elements exceeding the predefined phase-error threshold $\sigma_{th}$ are selectively disabled. Under this approach, RIS elements exceeding the predefined phase-error threshold $\sigma_{th},$ are disabled, reducing destructive interference by removing significantly misaligned element contributions. Consequently, deep nulls observed in the naive approach (Fig. \ref{fig:1}) become notably shallower, reflecting improved instantaneous received power in regions previously dominated by destructive interference. However, excluding RIS elements inherently reduces the total active reflective area, thereby diminishing coherent gain in regions of constructive interference. Hence, Fig. \ref{fig 2} clearly reveals the trade-off between mitigating interference and maintaining peak coherent power.
\subsection{Quantitative Evaluation of the Exclusion Method}
To quantitatively evaluate the effectiveness of the
proposed \textquotedbl Exclusion Method\textquotedbl{} defined in
\eqref{eq: Test} relative to the baseline \textquotedbl Naive Method\textquotedbl{}
(without excluding RIS elements in \eqref{eq:PR-B}), we introduce a quantitative performance metric $\Delta\bigl(\phi_{R},\sigma\bigr)$. This metric compares the expected received power under both scenarios by employing the logarithmic ratio:
\begin{equation}
\Delta\bigl(\phi_{R},\sigma\bigr)\triangleq10\log_{10}\left(\frac{\mathbb{E}\left[P_{\mathrm{R}}\bigl(\phi_{\mathrm{R}},\sigma_{th}\bigr)\right]}{\mathbb{E}\left[P_{\mathrm{R}}\bigl(\phi_{\mathrm{R}},\sigma\bigr)\right]}\right).
\end{equation}

Here, $\mathbb{E}\left[.\right]$ denotes the expectation over the uniform distribution of random phase errors. $\delta.$ A positive $\Delta\bigl(\phi_{R},\sigma\bigr)$ indicates that the exclusion strategy outperforms the naive method, enhancing the average received power at the phase shift $\phi_{R}$. When $\Delta\bigl(\phi_{R},\sigma\bigr)\thickapprox0$, both methods yield nearly equivalent performance, which typically occurs if very few RIS elements are excluded (small $\sigma$ or high threshold $\sigma_{th}$ ). Conversely, negative $\Delta\bigl(\phi_{R},\sigma\bigr)$ values imply that excluding RIS elements reduces the overall coherent gain due to fewer active reflective elements.

The expected value of the received power is mathematically derived as follows:
\begin{gather}
\mathbb{E}\left[P_{\mathrm{R}}\left(\phi_{\mathrm{R}},\delta\right)\right]=\nonumber \\
P_{\mathrm{t}}\left(A_{\mathrm{SU}}^{2}+A_{\mathrm{RU}}^{2}+2A_{\mathrm{SU}}A_{\mathrm{RU}}\:\mathbb{E}\left[\cos\left(\phi_{\mathrm{R}}-\Theta+\delta\right)\right]\right),\label{eq:PR(delta)-1}
\end{gather}
Applying the Law of the Unconscious Statistician (LOTUS), the critical expectation term is expressed as:
\begin{equation}
\mathbb{E}\left[\cos\bigl(\phi_{\mathrm{R}}-\Theta+\delta\bigr)\right]=\frac{1}{2\sigma}\intop_{-\sigma}^{\sigma}\cos\bigl(\phi_{\mathrm{R}}-\Theta+\delta\bigr)\:d\delta.\label{ED-B-1}
\end{equation}

Evaluating the integral yields a closed-form solution: 
\begin{gather}
\mathbb{E}\left[\cos\bigl(\phi_{\mathrm{R}}-\Theta+\delta\bigr)\right]=\mathrm{sinc}(\sigma)\cos(\phi_{\mathrm{R}}-\Theta)\label{eq: 24-1}
\end{gather}
with the $\mathrm{sinc}$ function defined explicitly as follows:

\begin{equation}
\text{sinc}\,(\sigma)\equiv\left\{ \begin{array}{cc}
1 & \text{ for }\sigma=0\\
\frac{\sin(\sigma)}{\sigma} & \text{ otherwise,}
\end{array}\right.
\end{equation}

Substituting \eqref{eq: 24-1} into \eqref{eq:PR(delta)-1}, the average
received power is expressed as follows:
\begin{gather}
\mathbb{E}\left[P_{\mathrm{R}}\left(\phi_{\mathrm{R}},\sigma\right)\right]=\nonumber \\
P_{\mathrm{t}}\left(A_{\mathrm{SU}}^{2}+A_{\mathrm{RU}}^{2}+2A_{\mathrm{SU}}A_{\mathrm{RU}}\:\mathrm{sinc}(\sigma)\cos(\phi_{\mathrm{R}}-\Theta)\right),\label{eq:PR-B-1}
\end{gather}

This formulation clearly illustrates that the term $\text{sinc}\,\sigma,$ represents the coherent gain reduction due to random phase errors.

\begin{figure}
\centering{}
\includegraphics[width=3.5in]{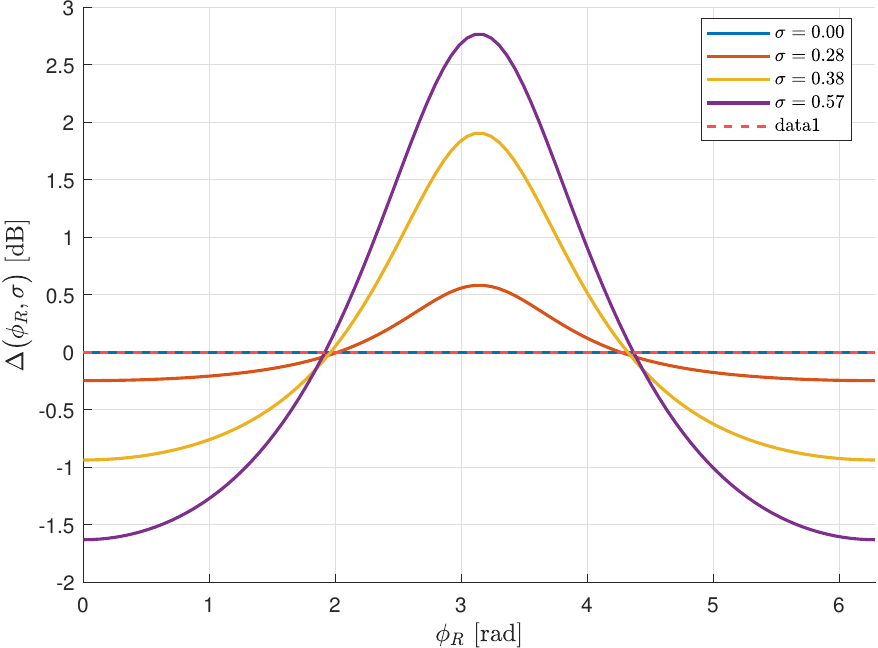}
\caption{Relative power difference due to phase errors and exclusion\label{fig 2B}}
\end{figure}

Fig. \ref{fig 2B} provides a quantitative evaluation by plotting $\Delta\bigl(\phi_{R},\sigma\bigr)$ across a range of RIS phase shifts. Positive $\Delta\bigl(\phi_{R},\sigma\bigr)$ values commonly occur near $\phi_{\mathrm{R}}\approx\pi$, a region where the naive method suffers from severe destructive interference. In contrast, selective exclusion enhances power by mitigating destructive interference. Negative values generally arise around regions of ideal constructive interference ($\phi_{\mathrm{R}}\approx0$, $2\pi$), indicating that the exclusion of RIS elements reduces the peak coherent gain.

 Collectively, the quantitative analysis demonstrated in Fig. \ref{fig:1}, \ref{fig 2}, and Fig. \ref{fig 2B} underscores the intricate trade-off between reducing destructive interference by excluding RIS elements and maintaining maximum coherent gain. These results highlight the critical importance of carefully selecting the exclusion threshold $\sigma_{th}$ to effectively balance interference mitigation and overall system performance.
\subsection{Joint Optimization of RIS Phase Shift and Exclusion Threshold}

Following the selective exclusion strategy for error-prone RIS elements, our next step involves jointly optimizing the RIS phase shift,  $\phi_{R}$, and the exclusion threshold, $\sigma_{th}$, to maximize the instantaneous received power $P_{R}$. This optimization problem is formally expressed as: 
\begin{gather}
\max_{\phi_{R},\sigma_{th}}P_{R}(\phi_{R},\sigma_{th})\nonumber \\
\textrm{subject to }\phi_{R}\in[0,2\pi],\:\sigma_{th}\in[\sigma_{th}^{min},\sigma_{th}^{max}].\label{eq: P_max}
\end{gather}
Here, the constraint $\phi_{R}\in[0,2\pi]$ ensures full-range phase control of the RIS elements, while $\sigma_{th}\in[\sigma_{th}^{min},\sigma_{th}^{max}]$ restricts the exclusion threshold within practical hardware and operational constraints.

Solving this joint optimization requires simultaneously adjusting $\phi_{R}$ and $\sigma_{th}$ to coherently align RIS-reflected signals with the direct path.This approach inherently balances the trade-off between aperture size (the number of active RIS elements) and phase coherence. Due to the intrinsic non-convexity and high computational complexity of this problem, Bayesian Optimization (BO) \cite{Gong2023}, a sample-efficient global optimization method, is utilized. 

BO employs a Gaussian Process (GP) surrogate model, denoted as $P_{R}(\phi_{R},\sigma_{th})\approx\mathcal{GP}\left(\mu,\mathcal{K}\right),$ characterized by a mean function $\mu(\phi_{R},\sigma_{\text{th}})$ and a covariance kernel $\mathcal{K}$. This kernel quantifies the similarity between any two parameter pairs in the optimization space, namely $(\phi_{R},\sigma_{\text{th}})$ and $(\phi_{R}^{'},\sigma_{\text{th}}^{'}),$ where $(\phi_{R}^{'},\sigma_{\text{th}}^{'})$ represents an arbitrary reference or comparison point within the parameter space. The covariance kernel is explicitly expressed as:
\begin{equation}
\mathcal{K}\left((\phi_{R},\sigma_{\text{th}}),(\phi_{R}^{'},\sigma_{\text{th}}^{'})\right)=\sigma_{f}^{2}\exp^{\left(-\frac{1}{2}\left[\frac{\bigl(\phi_{R}-\phi_{R}^{'}\bigr)^{2}}{l_{\phi}^{2}}+\frac{\bigl(\sigma_{\text{th}}-\sigma_{\text{th}}^{'}\bigr)^{2}}{l_{\sigma}^{2}}\right]\right)},
\end{equation}
where $\sigma_{f}^{2}$ denotes the signal variance, controlling the amplitude of variations in $P_{R}$ across the parameter domain, and $l_{\phi}$ and $l_{\sigma}$ are the length-scale parameters, governing how smoothly or rapidly $P_{R}$ changes with respect to $\phi_{R}$ and $\sigma_{th}$, respectively. Shorter length-scale values indicate rapid variations, whereas longer values correspond to smoother changes in the received power.

Central to the BO method is the expected improvement (EI) criterion, a quantitative measure used to assess the potential benefit of evaluating $P_{R}$ at new candidate parameter points $(\phi_{R},\sigma_{\text{th}})$. EI strategically balances exploration (evaluating uncertain regions of the parameter space) with exploitation (focusing on regions predicted to yield high levels of received power). Mathematically, EI is defined as:
\begin{equation}
EI(\phi_{R},\sigma_{\text{th}})=\left(\mu(\phi_{R},\sigma_{\text{th}})-f(x^{+})-\xi\right)\Phi(Z)+\sigma(\phi_{R},\sigma_{\text{th}})\phi(Z),
\end{equation}
where $\mu(\phi_{R},\sigma_{\text{th}})$ and $\sigma(\phi_{R},\sigma_{\text{th}})$
represent the GP-predicted mean and standard deviation of $P_{R}$
, $f(x^{+})$ is the highest currently observed value of $P_{R}$, and $\xi$ ($\xi\:\ge\:0$ ) is the parameter controlling the exploration-exploitation trade-off. A larger $\xi$ encourages exploration, while $\xi=0$ emphasizes exploitation. The standardized improvement score is 
$Z$ =$\frac{\mu(\phi_{R},\sigma_{\text{th}})-f(x^{+})}{\sigma(\phi_{R},\sigma_{\text{th}})},$ and $\Phi(Z)$ and $\phi(Z)$ denote the cumulative distribution function (CDF) and PDF of a standard normal distribution, respectively.

The optimization procedure commences with an initial evaluation of $P_{R}$ at selected parameter pairs $(\phi_{R},\sigma_{\text{th}}),$ creating the basis for the GP surrogate model. In subsequent iterations, the EI criterion is employed to identify promising new parameter combinations. As shown in Algorithm 1, this iterative process systematically explores the parameter space, prioritizing regions with either high predicted performance or significant uncertainty. It ultimately converges to the optimal combination that maximizes the instantaneous received power.
\begin{algorithm}[t]
\caption{RIS Phase Error Optimization with Bayesian Optimization}
\label{alg:RIS_BO}
\DontPrintSemicolon
\KwIn{
Phase shift range: $\phi_{\mathrm{R}} \in [0, 2\pi]$; Threshold bounds: $\sigma_{\text{th}} \in [\sigma_{\text{th}}^{\min}, \sigma_{\text{th}}^{\max}]$; \newline
Phase error distribution: $\delta_i \sim \mathcal{U}(-\sigma, \sigma)$; Initial samples: $N_{\text{init}}$; \newline
Maximum iterations: $T_{\text{max}}$; Exploration parameter: $\xi \geq 0$
}
\KwOut{Optimal parameters: $\phi_{\mathrm{R}}^*, \sigma_{\text{th}}^*$}

\textbf{Initialize:}
Generate random pairs $\{(\phi_{\mathrm{R}}^{(i)}, \sigma_{\text{th}}^{(i)})\}_{i=1}^{N_{\text{init}}}$

 \For{$i = 1$ \textbf{to} $N_{\text{init}}$}{
\For{$k = 1$ \textbf{to} $N$}{
Compute $\mathcal{S}_k = \mathbf{1}_{\{|\delta_k| > \sigma_{\text{th}}^{(i)}\}}$
}}
Evaluate
\small$P_{\mathrm{R}}^{(i)} = P_{\mathrm{t}}[A_{\mathrm{SU}}^2 + \sum_{k=1}^N (1-\mathcal{S}_k)A_{\mathrm{RU}_k}^2 + 2A_{\mathrm{SU}}\sum_{k=1}^N (1-\mathcal{S}_k)A_{\mathrm{RU}_k}\cos(\phi_{\mathrm{R}}^{(i)} - \Theta + \delta_k)]$
 
$f(x^+) = \max\{P_{\mathrm{R}}^{(i)}\}_{i=1}^{N_{\text{init}}}$

\For{$t = 1$ \textbf{to} $T_{\text{max}}$}{
Train GP: $(\mu,\sigma) \leftarrow \texttt{GP}(\mathcal{D}_{t-1})$

Compute EI:
$\text{EI} = (\mu - f(x^+) - \xi)\Phi(Z) + \sigma\phi(Z), \quad Z = \frac{\mu - f(x^+) - \xi}{\sigma}$

Select: $(\phi_{\mathrm{R}}^{(t)}, \sigma_{\text{th}}^{(t)}) = \arg\max \text{EI}$

Evaluate:
\small$P_{\mathrm{R}}^{(t)} = P_{\mathrm{t}}[A_{\mathrm{SU}}^2 + \sum_{k=1}^N (1-\mathcal{S}_k)A_{\mathrm{RU}_k}^2 + 2A_{\mathrm{SU}}\sum_{k=1}^N (1-\mathcal{S}_k)A_{\mathrm{RU}_k}\cos(\phi_{\mathrm{R}}^{(t)} - \Theta + \delta_k)]$

Update: $\mathcal{D}_t = \mathcal{D}_{t-1} \cup \{(\phi_{\mathrm{R}}^{(t)}, \sigma_{\text{th}}^{(t)}), P_{\mathrm{R}}^{(t)}\}$

\If{$P_{\mathrm{R}}^{(t)} > f(x^+)$}{
$f(x^+) \leftarrow P_{\mathrm{R}}^{(t)}$; $\phi_{\mathrm{R}}^* \leftarrow \phi_{\mathrm{R}}^{(t)}$, $\sigma_{\text{th}}^* \leftarrow \sigma_{\text{th}}^{(t)}$
}

\If{Converged ($\Delta P_{\mathrm{R}} < \epsilon$ for 5 iterations)}{
\textbf{break}
}
}
\Return $\phi_{\mathrm{R}}^*, \sigma_{\text{th}}^*$
\end{algorithm}

\begin{figure}
\centering{}
\includegraphics[width=3.5in]{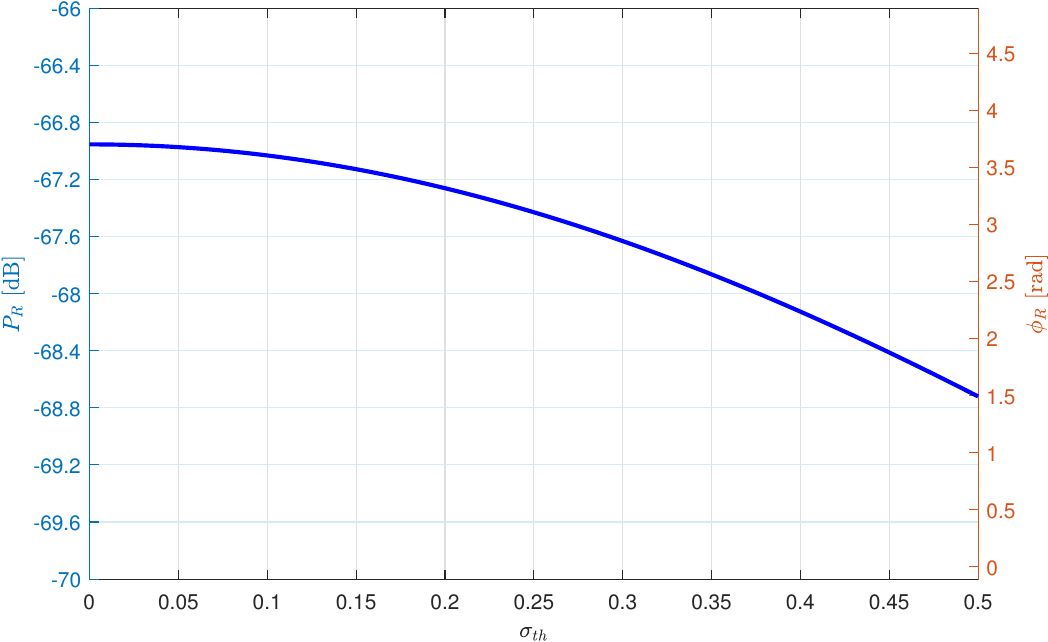}
\caption{Balancing the phase tuning accuracy and error resilience for enhanced
signal power in RIS-assisted satellite-to-Earth communication systems
\label{fig 3} }
\end{figure}

Fig. \ref{fig 3} validates the proposed optimization framework by illustrating the joint impact of the phase shift $\phi_{R}$ and exclusion threshold $\sigma_{th}$ on the received power $P_{R}$, utilizing parameters from Tabel-\ref{tab:Table-1}. The results indicate the presence of a critical operational threshold at $\sigma_{th}=0.13$, beyond which $P_{R}$ experiences significant degradation due to the exclusion of error-prone RIS elements. For $\sigma_{th}\:\le\:0.13$, performance remains stable at peak levels, suggesting negligible degradation from residual phase errors. The observed decline for $\sigma_{th}>0.13$ can be attributed to the systematic deactivation of RIS elements that exceed this phase error threshold, which effectively reduces the reflective aperture and disrupts coherent signal superposition.

\begin{figure}
\centering{}
\includegraphics[width=3.5in]{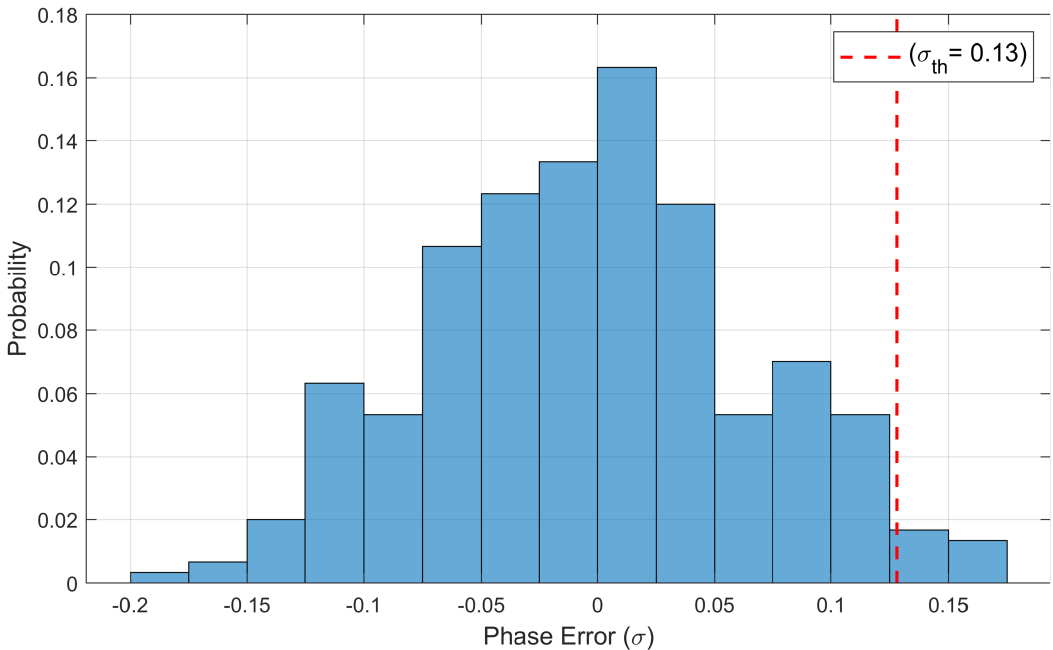}
\caption{Distribution of phase errors across RIS elements: statistical insights
and threshold optimization\label{fig 4} }
\end{figure}

To clarify the selected threshold, Fig. \ref{fig 4} demonstrates the probability distribution of phase errors $(\sigma)$ across RIS elements. The histogram shows that the phase errors vary between -0.2 and +0.2, with a critical threshold set at $\sigma_{th}=0.13$. Analysis of the data indicates that approximately 95\% of the phase error values fall within the range of $\sigma_{th}\:\le\:0.13$, while about 5\% of the values exceed this limit (i.e., $|\sigma|>0.13$). These outlier errors are more likely to introduce destructive interference into the system. Therefore, by selecting $\sigma_{th}=0.13$, we effectively exclude only the most extreme errors. Moreover, the presence of an asymmetrical distribution beyond the threshold highlights the necessity of precisely calibrating $\sigma_{th}$ to mitigate performance degradation caused by these outlier errors.

\begin{figure}
\centering{}\includegraphics[width=3.5in]{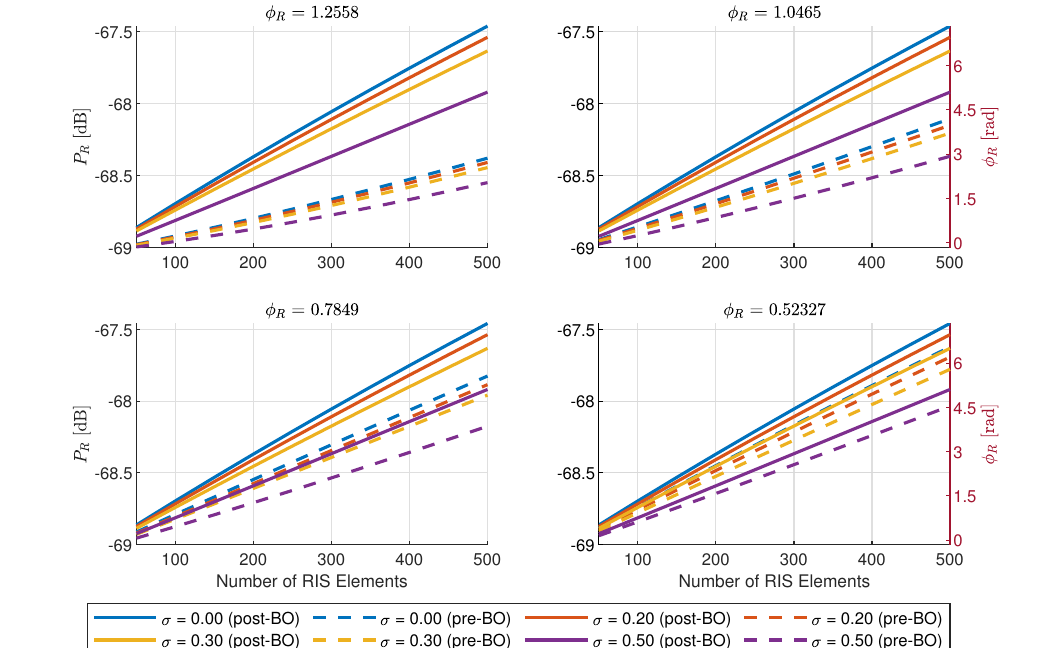} \caption{ Received signal power for different numbers of RIS
elements with and without BO under phase errors \label{fig 5} }
\end{figure}

 Fig. \ref{fig 5} demonstrates the impact of BO on enhancing the received power $P_{R}$ in a RIS-assisted satellite communication system. The figure compares two scenarios: a baseline  Pre-BO state and an optimized  Post-BO state.

In the Pre-BO scenario (dashed lines), the system operates with a fixed phase shift $\phi_{R}$ and a threshold $\sigma_{th}$=0.2 rad. RIS elements with phase errors exceeding this threshold are excluded, thereby establishing a performance baseline. Variations in $P_{R}$ with an increasing number of RIS elements reflect degradation due to high phase
errors.

In the Post-BO scenario (solid lines), BO optimizes both $\phi_{R}$ and $\sigma_{th}$ for each $\sigma$ setting, resulting in a notable increase in $P_{R}$ (an enhancement of 6.6\% and 9.2\% over the evaluated range). This improvement arises from improved phase alignment and the selective exclusion of elements with excessive errors.

Both scenarios show a steady increase in $P_{R}$ with more RIS elements; however, the Post-BO state consistently achieves higher power, demonstrating the effectiveness of BO in managing phase errors. Fig. \ref{fig 5} considers up to 500 RIS elements; it is worth noting that scaling to larger arrays will increase the computational complexity of BO, particularly in GP modeling. Future implementations may benefit from hierarchical optimization by dividing the RIS into smaller sub-arrays for independent optimization to reduce both the dimensionality and computational load.
\section{RIS Optimization Under Phase Errors and Shadowing Effects \label{sec:non_idel} }
In our previous analysis, we considered only phase errors in RIS elements to establish a clear theoretical benchmark, intentionally omitting
factors such as shadowing. However, for practical satellite-to-Earth communications in urban settings, it is essential to incorporate both
shadowing and atmospheric effects for accurate modeling. Since advanced technologies mitigate atmospheric effects (as noted in Section \ref{sec:Syst_Mod}), our focus now shifts to shadowing, which significantly influences signal transmission and reception in dense urban areas. In particular, we explore how shadowing interacts with RIS phase errors. Accurate modeling of these effects enables the design of RIS configurations and deployment strategies that are resilient to environmental challenges, thereby enhancing communication reliability.

Shadowing, caused by obstructions such as buildings and trees, introduces random signal attenuation that weakens both the direct satellite link and RIS-assisted paths. These effects are empirically modeled as log-normal distributed random variables with zero mean and standard deviations ranging from 6 dB to 12 dB, aligning with urban measurements \cite{Bertoni1999}. Furthermore, satellite LoS links experience additional shadow fading, which is formally recognized as an incremental path loss component in standards such as 3GPP \cite{3GPP2019}. 

The combined impact of shadowing and phase errors transforms the received power into a stochastic quantity. For the received power expression in \eqref{eq: Test} deterministic evaluation becomes infeasible due to the randomness of shadowing. Instead, statistical expectations must characterize system performance. This accounts for joint uncertainties arising from shadowing and phase errors, enabling realistic reliability assessments in urban environments. The expected received power is expressed as:
\begin{gather}
\mathbb{E}\left[P_{\mathrm{R}}(\phi_{\mathrm{R}},\sigma_{th})\right]=\mathbb{E}_{\delta,\eta_{SU},x_{RU}}\left[P_{\mathrm{t}}\left(A_{\mathrm{SU}}^{2}+\sum_{i=1}^{N}\textrm{\ensuremath{(1-\mathcal{S_{\mathit{i}}}}})A_{\mathrm{RU}_{i}}^{2}\right.\right.\nonumber \\
\left.\left.+2A_{\mathrm{SU}}\sum_{i=1}^{N}\textrm{\ensuremath{(1-\mathcal{S_{\mathit{i}}}}})A_{\mathrm{RU}_{i}}\cos\left(\phi_{R}^{(j)}-\Theta+\delta_{i}\right)\right)\right]\label{eq: GA}
\end{gather}
where $\mathbb{E}_{\delta,\eta_{SU},x_{RU}}$is the expectation over  the distributions of $\delta,\eta_{SU},$ and $x_{RU}.$

The expectation can be computed via integration over the joint PDFs of $\delta$, $\eta_{SU}$, and $x_{RU}$ as follows: 
\begin{gather}
\mathbb{E}\left[P_{\mathrm{R}}\bigl(\phi_{\mathrm{R}}\bigr)\right]=\int_{-\sigma}^{\sigma}\int_{-\infty}^{\infty}\int_{-\infty}^{\infty}P_{R}(\phi_{R},\delta,\eta_{SU},x_{RU})\nonumber \\
\:f_{\delta}(\delta)f_{\eta_{SU}}(\eta_{SU})f_{x_{RU}}(x_{RU})\,d\delta\,d\eta_{SU}\,dx_{RU}\label{eq:33}
\end{gather}
where $f_{\delta}(\delta),f_{\eta_{SU}}(\eta_{SU})$, and $f_{x_{RU}}(x_{RU})$ are the respective PDFs of the phase error and the shadowing processes associated with both the S-U and R-U links, respectively.

Due to the complexity of \eqref{eq:33}, the expectation is estimated
numerically using Monte Carlo methods. This method effectively captures
the random variability of phase errors through the generation of representative
sample sets of the phase error $\left\{ \delta_{1},\delta_{2},...,\delta_{K}\right\} $,
S-U link shadowing $\left\{ \eta_{\mathrm{SU_{1}}},\eta_{\mathrm{SU_{2}}},...,\eta_{\mathrm{SU}_{K}}\right\} $
and R-U channel shadowing $\left\{ x_{\mathrm{RU_{1}}},x_{\mathrm{RU_{2}}},...,x_{\mathrm{RU}_{K}}\right\} $.
For each sample $k$, the received power is calculated for a specified RIS phase shift $\phi_{R}^{(j)}$, as follows: 
\begin{gather}
\mathbb{E}\left[P_{\mathrm{R}}\left(\phi_{R}^{(j)},\sigma_{th}\right)\right]\approx\frac{1}{K}\sum_{k=1}^{K}\left[P_{\mathrm{t}}\left(A_{\mathrm{SU}_{k}}^{2}+\sum_{i=1}^{N}1-\mathcal{S}_{i}(\delta_{i},\sigma_{th})A_{\mathrm{RU}_{i,k}}^{2}\right.\right.\nonumber \\
\left.\left.+2A_{\mathrm{SU}_{k}}\sum_{i=1}^{N}1-\mathcal{S}_{i}(\delta_{i},\sigma_{th})A_{\mathrm{RU}_{i,k}}\cos\left(\phi_{R}^{(j)}-\Theta+\delta_{i,k}\right)\right)\right]\label{eq: GA-1}
\end{gather}
where $\mathcal{S}_{i}(\delta_{i},\sigma_{th})$ is a threshold function
that deactivates the $i^{th}$ RIS element if $\left|\delta_{i,k}\right|>\sigma_{th}$,
$A_{\mathrm{SU}_{k}}$ and $A_{\mathrm{RU}_{i,k}}$ are amplitudes that depend on shadowing samples $\eta_{SU_{k}}$, and $x_{RU_{i,k}}$, respectively. 
\begin{figure}
\centering{}
\includegraphics[width=3.5in]{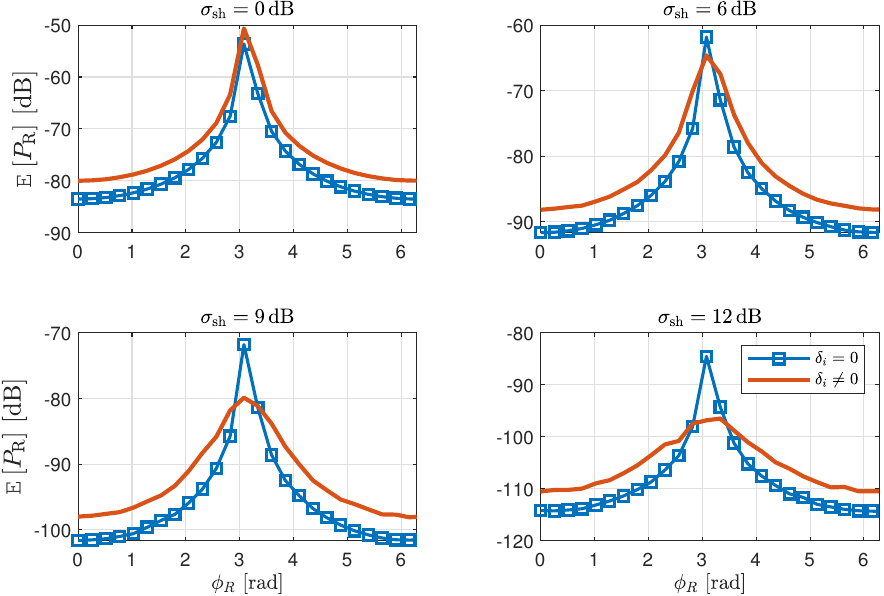}
\caption{The influence of shadowing and phase error on received power \label{fig 6} }
\end{figure}
Fig. \ref{fig 6} illustrates the expected received power as a function of the phase shift setting across various shadowing levels, both in the absence and presence of phase errors. The shadowing levels, defined
at 0 dB, 6 dB, 9 dB, and 12 dB, emulate environmental conditions
typical of urban areas, where obstacles such as buildings and foliage can impede signal propagation. In each plot, one curve, rendered in blue, represents scenarios without phase errors and, therefore, exhibits
optimal performance, while a second curve, shown in orange, reflects
the influence of phase errors, resulting in a reduction of received power. This reduction becomes more pronounced as the shadowing level
increases, thereby emphasizing the compounded effect of environmental
factors and phase errors on system performance. The peaks of these
curves denote the optimal phase shift settings that maximize received power, underscoring the importance of precise phase alignment in mitigating
environmental challenges in satellite-to-Earth communication systems.

To effectively optimize the expected received power in the presence of shadowing and phase errors, our strategy is to identify the optimal settings for $\phi_{\mathrm{R}}$ and $\sigma_{th}$ that maximize the expected power. The optimization objective is formally defined as:
\begin{equation}
\overset{*}{\phi_{R}},\overset{*}{\sigma_{th}}=\mathrm{arg\quad}\max_{\underset{\underset{\mathrm{min}}{\sigma}\leq\sigma_{th}\leq\underset{\mathrm{max}}{\sigma}}{0\leq\phi_{R}\leq2\pi}}\mathbb{E}[P_{R}(\phi_{R},\sigma_{th})],\label{eq: opt_Shwd}
\end{equation}
where the optimization ensures that all parameter combinations remain within their respective feasible regions.

To address this stochastic optimization problem, we employ BO with a GP as the Surrogate model to approximate the expected received power $\mathbb{E}[P_{R}(\phi_{R},\sigma_{th})]$. The GP is characterized by its mean function $m(\phi_{R},\sigma_{th})$ and the covariance function $\mathcal{K}((\phi_{R},\sigma_{th}),(\phi_{R}',\sigma_{th}'))$.
The optimization process begins with the construction of an initial dataset $\mathcal{D}$ consisting of $N_{\mathrm{int}}$ points. These points are sampled uniformly from the parameter space defined by the allowable ranges for $\phi_{R}$ and $\sigma_{th}$, and the expected received power is evaluated at each point, yielding
\begin{equation}
\mathcal{D}=\left\{ (\phi_{R}^{(i)},\sigma_{th}^{(i)},y^{(i)})\right\} _{i=1}^{N_{\text{int}}},\textrm{where \ensuremath{y^{(i)}}=\ensuremath{\mathbb{E}}[\ensuremath{P_{R}}(\ensuremath{\phi_{R}^{(i)}},\ensuremath{\sigma_{th}^{(i)}})].}
\end{equation}

At each subsequent iteration $k$, a new point is selected according to the chosen acquisition function, evaluated, and this new observation is added to the dataset. Consequently, after
$k$ iterations, the dataset is updated as $\mathcal{D}=\left\{ (\phi_{R}^{(i)},\sigma_{th}^{(i)},y^{(i)})\right\} _{i=1}^{N_{\text{int}}+k}.$

Subsequent sampling points are selected through a dual-phase acquisition strategy that alternates between the upper confidence bound (UCB) and expected improvement (EI) criteria. This hybrid approach balances the exploration of uncertain regions and the exploitation of promising areas. At each iteration $k+1$, the next evaluation point is determined by maximizing the acquisition function $a(\phi_{R},\sigma_{th}|\mathcal{D})$:
\begin{equation}
\biggl(\overset{\bigl(k+1\bigr)}{\phi_{R}},\overset{\bigl(k+1\bigr)}{\sigma_{th}}\biggr)=\mathrm{arg}\max_{\underset{\underset{\mathrm{min}}{\sigma}\leq\sigma_{th}\leq\underset{\mathrm{max}}{\sigma}}{0\leq\phi_{R}\leq2\pi}}a(\phi_{R},\sigma_{th}|\mathcal{D}).
\end{equation}
During initial iterations, UCB is prioritized to encourage global exploration. The UCB acquisition function incorporates both the GP posterior mean $m(\phi_{R},\sigma_{th})$ and uncertainty $\sigma(\phi_{R},\sigma_{th})$, expressed as 
\begin{equation}
\mathrm{UCB}(\phi_{R},\sigma_{th})=m(\phi_{R},\sigma_{th})+\kappa\cdot\sigma(\phi_{R},\sigma_{th}),
\end{equation}
where $\kappa>0$ modulates the exploration-exploitation trade-off.

As the optimization progresses and high-potential regions are identified, the strategy transitions to EI for localized refinement. The EI criterion quantifies the expected improvement over the current optimum $y_{best}=\underset{1\leq i\leq k}{\mathrm{max}}\mathbb{E}[P_{R}(\phi_{R}^{(i)},\sigma_{th}^{(i)})]$
\begin{equation}
\mathrm{EI}(\phi_{R},\sigma_{th})=\mathbb{E}\bigl[\max\,\bigl(0,y-y_{\text{best}}\bigr)\bigr],
\end{equation}

After evaluating the selected point $\left(\phi_{R}^{(k+1)},\sigma_{th}^{(k+1)}\right),$ the dataset $\mathcal{D}$ is expanded by adding the new observation $\left(\phi_{R}^{(k+1)},\sigma_{th}^{(k+1)},y^{(k+1)}\right)$ according to
\begin{equation}
\mathcal{D}=\mathcal{D}\cup\left\{ \left(\phi_{R}^{(k+1)},\sigma_{th}^{(k+1)},y^{(k+1)}\right)\right\} 
\end{equation}
where $\ensuremath{\bigcup}$ is the union operation, denoting the addition of a new observation to the existing dataset. 

This iterative process repeats until the convergence criteria (e.g., maximum iterations or negligible improvement) are met, yielding the optimal parameters $\Bigl(\overset{*}{\phi_{R}},\overset{*}{\sigma_{th}}\Bigr)$ that maximize the expected received power while mitigating the combined effects of shadowing and phase errors. This process is detailed in Algorithm 2.
\begin{algorithm} 
\caption{Bayesian Optimization for Received Power Maximization}
\label{alg:BO_optimization}
\KwIn{Phase shift bounds: \(0 \leq \phi_R \leq 2\pi\)\\
Threshold deviation bounds: \(\sigma_{\text{min}} \leq \sigma_{th} \leq \sigma_{\text{max}}\)\\
Initial dataset size: \(N_{\text{int}}\)\\
Acquisition parameters: \(\kappa\) (UCB weight), \(y_{\text{best}}\) (initial best value)\\
GP prior: Mean \(m(\cdot)\), kernel \(\mathcal{K}(\cdot, \cdot)\)\\
Stopping criteria: \(K_{\text{max}}\) (max iterations), \(\epsilon\) (tolerance)}
\KwOut{Optimal parameters: \(\left(\phi_{R}^*, \sigma_{th}^*\right)\)}

\BlankLine
Initialize \(\mathcal{D}\) with \(N_{\text{int}}\) points:\\
\For{\(i = 1\) to \(N_{\text{int}}\)}{
    Sample \(\phi_R^{(i)} \sim \mathcal{U}(0, 2\pi)\), \(\sigma_{th}^{(i)} \sim \mathcal{U}(\sigma_{\text{min}}, \sigma_{\text{max}})\)\\
    Compute \(y^{(i)} = \mathbb{E}[P_R(\phi_R^{(i)}, \sigma_{th}^{(i)})]\)\\
    \(\mathcal{D} \leftarrow \mathcal{D} \cup \left\{ \left(\phi_R^{(i)}, \sigma_{th}^{(i)}, y^{(i)}\right) \right\}\)
}

\For{\(k = 1\) to \(K_{\text{max}}\)}{
    Train GP on \(\mathcal{D}\) to update \(m(\cdot)\) and \(\mathcal{K}(\cdot, \cdot)\)\\
    \eIf{\(k \leq K_{\text{switch}}\)}{
        Select \(\left(\phi_R^{(k+1)}, \sigma_{th}^{(k+1)}\right)\) by maximizing \(\text{UCB}(\phi_R, \sigma_{th})\)\\
        \(\text{UCB}(\phi_R, \sigma_{th}) = m(\phi_R, \sigma_{th}) + \kappa \cdot \sigma(\phi_R, \sigma_{th})\)
    }{
        Select \(\left(\phi_R^{(k+1)}, \sigma_{th}^{(k+1)}\right)\) by maximizing \(\text{EI}(\phi_R, \sigma_{th})\)\\
        \(\text{EI}(\phi_R, \sigma_{th}) = \mathbb{E}\left[\max\left(0, y - y_{\text{best}}\right)\right]\)
    }
    Compute \(y^{(k+1)} = \mathbb{E}[P_R(\phi_R^{(k+1)}, \sigma_{th}^{(k+1)})]\)\\
    Update \(\mathcal{D} = \mathcal{D} \cup \left\{ \left(\phi_R^{(k+1)}, \sigma_{th}^{(k+1)}, y^{(k+1)}\right) \right\}\)\\
    Update \(y_{\text{best}} = \max\left(y_{\text{best}}, y^{(k+1)}\right)\)\\
    \If{\(\left|y^{(k+1)} - y_{\text{best}}\right| < \epsilon\)}{
        \textbf{break}
    }
}
\Return \(\left(\phi_{R}^*, \sigma_{th}^*\right) = \underset{\mathcal{D}}{\arg\max}\, y^{(i)}\)
\end{algorithm}

 To evaluate the practicality of our optimization approach, we examine its computational complexity. The proposed framework involves two main procedures: evaluating the received power $P_{R}$, which scales linearly with the number of RIS elements $N$ and Monte Carlo samples $K$, and exploring the parameter space using BO with a GP surrogate. Updating the GP model at each iteration requires $\mathcal{O}(M^{3})$ complexity, where $M$ is the cumulative number of function evaluations. Thus, the total complexity grows as $\mathcal{O}(M^{3}+MNK)$. 

 While this is higher than simpler approaches that neglect phase errors or shadowing, the complexity remains manageable due to efficient BO sampling and the linear, parallelizable calculations over RIS elements. Therefore, the overall computational burden scales predictably and remains feasible for realistic satellite-to-ground applications. 

\begin{figure}
\centering{}
\includegraphics[width=3.5in]{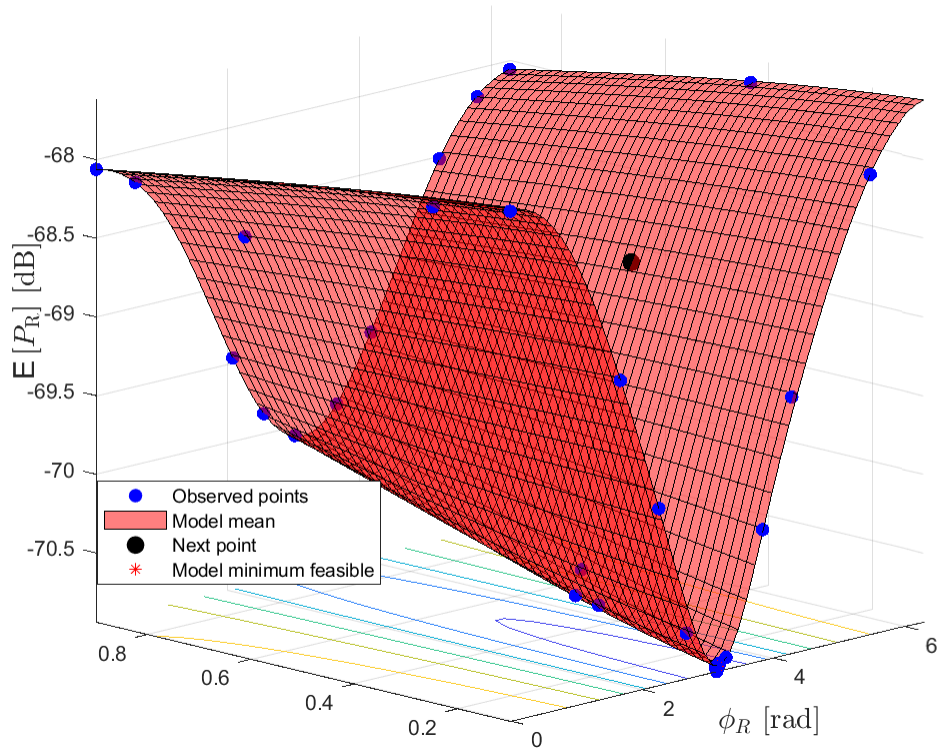}
\caption{3D visualization of optimization objective function values through
$\phi_{\mathrm{R}}$ and $\sigma_{th}$\label{fig 7} }
\end{figure}

The 3D surface in Fig. \ref{fig 7} represents the estimated expected received power $\mathbb{E}[P_{R}]$ (in dB) as a function of the $\phi_{\mathrm{R}}$ and $\sigma_{th},$, providing insights into how Algorithm-2 explores the parameter space. This visualization highlights optimal regions where $\mathbb{E}[P_{R}]$ is maximized, with peaks indicating favorable settings and valleys representing less effective configurations.

Markers on the surface denote \textit{Observed Points}, which represent the actual evaluations of $\mathbb{E}[P_{R}]$ at specific parameter pairs $(\phi_{R}^{(i)},\sigma_{th}^{(i)})$. These points help refine the GP model, improving its accuracy and enabling it to make more reliable predictions for unexplored regions. The surface itself represents the \textit{Model Mean}, which integrates observed data with the GP's probabilistic trends, providing a smooth interpolation of known results and an extrapolation into unknown areas. The surface also highlights the \textit{Next Point}, $\biggl(\overset{\bigl(k+1\bigr)}{\phi_{R}},\overset{\bigl(k+1\bigr)}{\sigma_{th}}\biggr)$, selected by the acquisition function. This point reflects the algorithm's strategy of balancing exploration of uncertain regions with exploitation of high-performance areas near known peaks. 

The \textit{Model Minimum} pinpoints the lowest predicted $\mathbb{E}[P_{R}]$ within feasible bounds, steering the algorithm away from impractical configurations where shadowing or phase errors dominate. The clustering of observed points around optimal peaks demonstrates the algorithm's progression toward high-performing solutions, while the sharp decline in $\mathbb{E}[P_{R}]$ for larger values of $\sigma_{th}$ emphasizes the sensitivity of received power to shadowing effects. Overall, this visualization provides a clear picture of the interaction between data-driven modeling and informed parameter selection, underscoring the effectiveness of the optimization framework.

\begin{figure}
\centering{}
\includegraphics[width=3.5in]{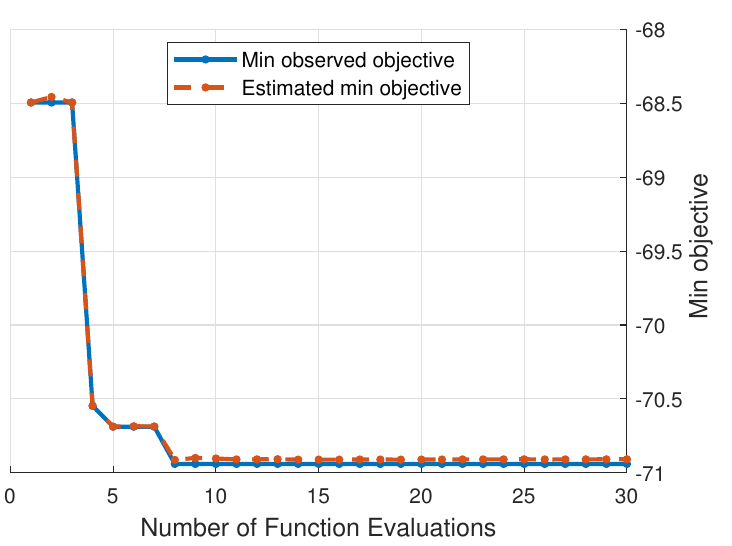}
\caption{Minimum and estimated minimum observed objective with respect to function
evaluation iterations \label{fig 8} }
\end{figure}

Fig \ref{fig 8} illustrates the convergence behavior of the BO process through two key metrics: the Minimum Observed Objective (lowest measured $\mathbb{E}[P_{R}]$ from actual evaluations) and the Estimated Minimum Objective (lowest predicted $\mathbb{E}[P_{R}]$ by the GP surrogate model). The vertical axis quantifies $\mathbb{E}[P_{R}]$ in dB, where lower (more negative) values correspond to weaker received power, while the horizontal axis tracks the progression of function evaluations (iterations).

The close alignment between the two curves validates the GP model's accuracy in approximating the true objective function. During the initial phase (first 5 evaluations), both metrics remain near \textminus 68 dB, reflecting the algorithm's exploration of the parameter space to identify promising regions. A sharp decline in both curves occurs at the $5^{th}$ evaluation, where $\mathbb{E}[P_{R}]$ drops to approximately -71 dB, signaling the discovery of an optimal configuration that balances phase alignment ($\phi_{\mathrm{R}}$) and shadowing mitigation ($\sigma_{th}$). Subsequent evaluations refine this solution, with both curves stabilizing near -71 dB, indicating convergence as further improvements diminish.

This stabilization suggests that the algorithm has identified a near-optimal parameter set, with additional evaluations yielding negligible improvements. The transition from exploration to exploitation aligns with the BO framework's strategic use of acquisition functions to prioritize high-confidence regions after initial sampling. The figure underscores the GP model's reliability in guiding the optimization process toward robust solutions under shadowing and phase error constraints.

\begin{figure}
\centering{}
\includegraphics[width=3.5in]{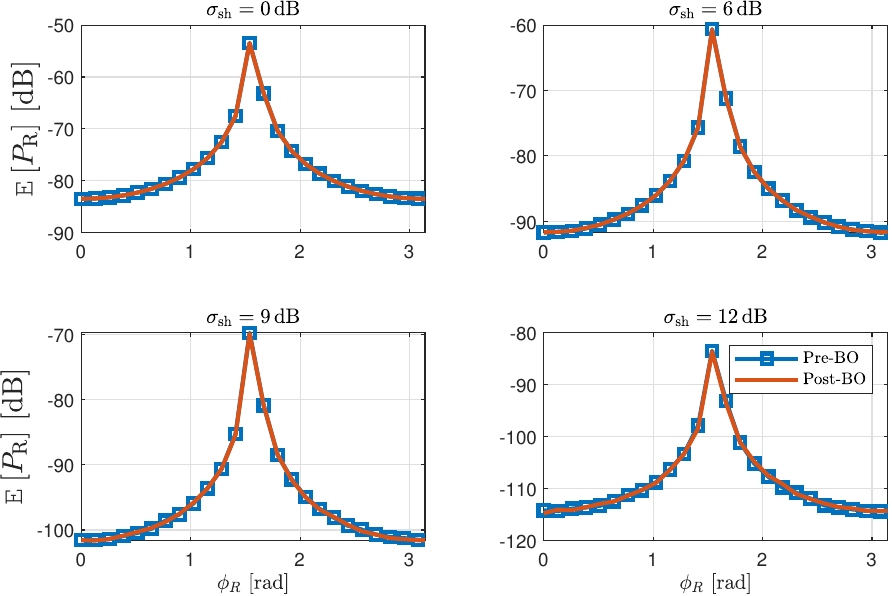}
\caption{Effectiveness of Bayesian optimization in mitigating phase errors
under shadowing effects\label{fig 9} }
\end{figure}

Fig \ref{fig 9} evaluates the performance of the BO framework in enhancing the expected received power, $\mathbb{E}[P_{R}]$, under varying shadowing conditions ($\sigma_{sh}=0,6,9,12$ dB). Each subplot corresponds to a specific \textbf{$\sigma_{sh}$} level, with the horizontal axis representing the phase shift $\phi_{\mathrm{R}}$ and the vertical axis quantifying $\mathbb{E}[P_{R}]$ in dB. The  Pre-BO curves depict baseline performance under ideal phase alignment (\textbf{$\delta_{i}=0$}, no phase errors), while the  Post-BO curves incorporate phase errors ($\delta_{i}\neq0$) but utilize optimized $\phi_{\mathrm{R}}$ configurations derived from the BO process.

The Post-BO results align closely with the  Pre-BO baseline across all shadowing levels, demonstrating the algorithm's ability to mitigate phase errors while maintaining robust performance under shadowing constraints. For example, at $\sigma_{sh}=0$ (minimal shadowing), the Post-BO curve achieves near-ideal performance, improving $\mathbb{E}[P_{R}]$ by 12.6\% compared to the  Pre-BO baseline. For moderate shadowing ($\sigma_{sh}=6$ dB), the optimization retains robustness, yielding a 10.2\% gain in $\mathbb{E}[P_{R}]$. Even under severe shadowing ($\sigma_{sh}=12$ dB), the framework maintains efficacy, delivering a 9.3\% improvement in $\mathbb{E}[P_{R}]$.

This consistency highlights the BO framework's dual capability to precisely calibrate $\phi_{\mathrm{R}}$ to counteract phase errors and strategically optimize under fixed shadowing constraints. The alignment of \textquotedblleft Post BO\textquotedblright{} results with \textquotedblleft Pre-BO\textquotedblright{} baselines, despite the introduction of practical impairments, validates the algorithm's ability to navigate parameter trade-offs while maintaining system performance. The improvements across $\sigma_{sh}$ levels emphasize the framework's utility in real-world scenarios, where shadowing is an inherent environmental factor rather than a variable to be mitigated.

\section{Conclusion and Future work \label{Conclusion} }
This paper presents a framework for optimizing received power in RIS-assisted satellite-to-Earth communications by accounting for both RIS hardware-induced phase errors and practical shadowing effects. A Bayesian optimization method was developed to dynamically configure RIS elements by excluding those that are prone to errors, thereby balancing the enhancement of signal power with the preservation of signal quality. Under conditions dominated by phase errors, the proposed approach yielded a 9.2\% improvement in received power, while incorporating shadowing effects further increased the enhancement to 12.6\%.

Future work will focus on scaling the framework to accommodate large RIS arrays, thereby enhancing signal manipulation
capabilities. Optimization algorithms will be refined to manage increased complexity, potentially integrating advanced machine learning methods to improve efficiency. This expansion is crucial for boosting the performance of communication systems employing extensive RIS infrastructures. Additionally, the research will address imperfect channel state information by refining the BO framework to enhance both the estimation and management of channel uncertainties, thereby increasing the system's resilience in practical deployments. Concurrent efforts will expand on the multiplicative error model, developing strategies to dynamically adjust for variations in error scale and mitigate their impact. Moreover, the analysis will be extended to include small-scale fading, broadening the current focus on free-space channel models. By incorporating these factors, the enhanced system model is expected to better reflect real-world conditions, offering robust and scalable solutions for next-generation communication networks.

\bibliographystyle{IEEEtran} 
\addcontentsline{toc}{section}{\refname}\bibliography{RMIT}

\end{document}